\documentclass[aps,pra,twocolumn,superscriptaddress]{revtex4-2}
\usepackage{amsfonts}
\usepackage{graphicx}
\usepackage{epstopdf}
\usepackage{epsfig}
\usepackage{amsmath}
\usepackage[normalem]{ulem}
\usepackage{multirow}
\usepackage{makecell}
\usepackage{array}
\usepackage{booktabs}
\usepackage{mathtools}
\usepackage{bm}
\usepackage{color}
\usepackage{soul,xcolor}
\usepackage{array}
\usepackage{booktabs}

\begin{document}
\setstcolor{red}

\title{Geometry-induced wavefunction collapse}

\author{Li-Li Ye}
\affiliation{Lanzhou Center for Theoretical Physics, Key Laboratory of Theoretical Physics of Gansu Province, and Key Laboratory for Magnetism and Magnetic Materials of MOE, Lanzhou University, Lanzhou, Gansu 730000, China}
\affiliation{School of Electrical, Computer and Energy Engineering, Arizona State University, Tempe, Arizona 85287, USA}

\author{Chen-Di Han}
\affiliation{School of Electrical, Computer and Energy Engineering, Arizona State University, Tempe, Arizona 85287, USA}

\author{Liang Huang} \email{huangl@lzu.edu.cn}
\affiliation{Lanzhou Center for Theoretical Physics, Key Laboratory of Theoretical Physics of Gansu Province, and Key Laboratory for Magnetism and Magnetic Materials of MOE, Lanzhou University, Lanzhou, Gansu 730000, China}

\author{Ying-Cheng Lai} \email{Ying-Cheng.Lai@asu.edu}
\affiliation{School of Electrical, Computer and Energy Engineering, Arizona State University, Tempe, Arizona 85287, USA}
\affiliation{Department of Physics, Arizona State University, Tempe, Arizona 85287, USA}

\date{\today}

\begin{abstract}

	When a quantum particle moves in a curved space, a geometric potential can arise. In spite of a long history of extensive theoretical studies, to experimentally observe the geometric potential remains to be a challenge. What are the physically observable consequences of such a geometric potential? Solving the Schr\"{o}dinger equation on a truncated conic surface, we uncover a class of quantum scattering states that bear a strong resemblance with the quasi-resonant states associated with atomic collapse about a Coulomb impurity, a remarkable quantum phenomenon in which an infinite number of quasi-resonant states emerge. A characteristic defining feature of such collapse states is the infinite oscillations of the local density of states (LDOS) about the zero energy point separating the scattering from the bound states. The emergence of such states in the curved (Riemannian) space requires neither a relativistic quantum mechanism nor any Coulomb impurity: they have zero angular momentum and their origin is purely geometrical - henceforth the term {\em geometry-induced} wavefunction collapse. We establish the collapsing nature of these states through a detailed comparative analysis of the behavior of the LDOS for both the zero and finite angular-momentum states as well as the corresponding classical picture. Potential experimental schemes to realize the geometry-induced collapse states are articulated. Not only has our study uncovered an intrinsic connection between the geometric potential and atomic collapse, it also provides a method to experimentally observe and characterize geometric potentials arising from different subfields of physics. For example, in nanoscience and nanotechnology, curved geometry has become increasingly common. Our finding suggests that wavefunction collapse should be an important factor of consideration in designing and developing nanodevices.

\end{abstract}

\date{\today}

\maketitle

\section{Introduction} \label{sec:intro}

When a quantum particle moves on a curved surface, a geometric potential can
arise~\cite{JK:1971,daCosta:1981}, which is fundamental to quantum mechanics 
in the Riemannian geometry. However, it remains a challenge to experimentally 
observe the geometric potential~\cite{NKKS:2001,SDHKNTL:2010,OISYK:2012}. The 
main message of this work is that the conic geometric potential can induce 
wavefunction collapse as manifested by the peculiar behavior of the local 
density of states (LDOS) typically seen in atomic collapse. 
Semiclassically, this geometry-induced collapse phenomenon is manifested as 
particle's spiraling inward towards a region of large curvature in the 
classical-quantum correspondence~\cite{shortley:1931,shytov2007atomic}. Thus,
theoretically, our work unveils a natural connection between the quantum
mechanics in the curved space and the phenomenon of atomic collapse.
Experimentally, our finding provides a viable way to meet the challenge 
of experimentally observing the geometric potential by putting forward 
measurable quantities as in the recent experimental study of atomic collapse.

The radial component of the Schr\"{o}dinger equation for a particle on a conic 
surface~\cite{FM:2008} can be simplified as the Bessel equation with the 
$1/r^2$ effective potential. Historically, the study of the $1/r^2$ potential 
in 3D has a long history~~\cite{nicholson:1962,LL:book,coon:2002,hammer:2006,shortley:1931,case:1950}, which can be induced by diverse physical mechanisms 
such as particle-charge interactions~\cite{coon:2002,nishida:202} and Efimov 
physics~\cite{efimov:1970,efimov:1973}. For example, in the early work by 
Shortley~\cite{shortley:1931} in 1932, the wavefunction was set to be zero at 
the origin. In the work of Case~\cite{case:1950} in 1950, a fixed phase was 
required for the wavefunctions at the origin. Bound and scattering states 
under the hard-core boundary condition and zero net outflow from the 
scattering region were analyzed earlier by Nicholson~\cite{nicholson:1962} in 
1962 and more recently by Coon et al.~\cite{coon:2002} in 2002. That the 3D 
central $1/r^2$ potential can induce a fall to the center associated with 
both bound and scattering states was 
analyzed~\cite{LL:book,perelomov:1970,alliluev:1972}. There were also works 
on the 3D central $1/r^2$ potential from different perspectives, such as 
anomalous symmetry breaking~\cite{coon:2002} and limit 
cycles~\cite{hammer:2006}. 

A recent development in quantum physics is the experimental observation of 
atomic collapse~\cite{Wangetal:2013}, a phenomenon that was predicted nearly 
eighty years ago~\cite{PS:1945,ZP:1972,Greiner:book} to occur in an atom with 
a super-heavy nucleus. In the present work, we consider particle motion on a 
curved surface that gives rise to a $1/r^2$ potential in 2D~\cite{daCosta:1981,FM:2008,PNLM:2010,FSA:2012,ST:2012,DWLKLZ:2016,jiang:2021}. 
The main contribution of our work is the establishment of the connection 
between the quantum behaviors on a curved surface and those associated with 
atomic collapse, providing a feasible way to experimentally observe the 
geometric potential. To place our work in a proper context and to better 
explain our finding, here we provide a brief description of the phenomenon 
of atomic collapse.

Consider a hydrogen-like atom of nuclear charge $Z$ with the Coulomb potential 
$-Z/r$. For $Z > 1/\alpha_0$, where $\alpha_0\equiv e^2/(\hbar c)\approx 1/137$
is the vacuum fine structure constant, the eigenenergy becomes complex, 
signifying the emergence of a resonant state with a finite lifetime for the 
electron that is inversely proportional to the imaginary part of the 
eigenenergy. The physical picture is that, in a sufficiently strong Coulomb 
field, the eigenenergy dives into the hole continuum, and the laws of 
relativistic quantum mechanics stipulate the creation of an electron-positron 
pair. Once this happens, the positron is free but the electron and the nucleus 
will form a quasi-bound resonant state, as if the electron had collapsed onto 
the nucleus. From a classical point of view, the electron behaves as if it 
spiraled inward toward the nucleus. Because of the finite lifetime of the 
resonant state, the electron will eventually escape the nucleus and couple to 
the positron~\cite{Boyer:2004}. The wavefunction thus contains two components: 
one around the Coulomb singularity and another extending to infinity.

From a mathematical point of view, the Dirac equation breaks down in the 
vicinity of the $1/r$ singularity of the Coulomb potential and some regularized
form of the potential should then be used so that the Dirac equation remains 
valid. Even then, for sufficiently large values of $Z$, the eigenenergies will 
still be complex. A general estimate of the required $Z$ values for atomic 
collapse to occur~\cite{PS:1945,alliluev:1972} is $Z > 170$, which exceeds the 
largest known atomic number of any natural element with the fine structure 
constant $\alpha_0$. To experimentally realize atomic collapse, some kind of 
relativistic quantum materials with a much larger effective fine-structure 
constant (or a much reduced ``speed of light'') can be exploited. In graphene, 
the Fermi velocity of the relativistic quantum quasiparticles is about two 
orders of magnitude smaller than the vacuum speed of light, so the effective 
fine-structure constant is on the order of unity, making possible experimental 
observation of atomic collapse~\cite{SKL:2007a,SKL:2007b}. This perspective 
stimulated theoretical studies of the various aspects of the energy states of 
an atomic impurity embedded in graphene such as 
screening~\cite{FNS:2007,TMKS:2008},
density of states~\cite{SKL:2007a,SKL:2007b,PNN:2007,NKNPPU:2009}, scattering
phase~\cite{SKL:2007b,Novikov:2007}, and generalization taking into account
electron-electron interactions~\cite{KUPGN:2012}. In 2012, the first
experimental observation of atomic collapse in graphene was
achieved~\cite{Wangetal:2012,Wangetal:2013}, generating subsequent interest
in this phenomenon~\cite{Maoetal:2016,jiang2017tuning,OMJAA:2017}. Quite
recently, atomic collapse has been predicted to occur in pseudospin-1 Dirac
materials with a flat band~\cite{HXHL:2019}.

The general feature of atomic collapse, i.e., the emergence of an infinitely
many resonant states~\cite{SKL:2007b}, can be understood by considering Dirac
fermions with energy $\varepsilon < 0$ in the two-dimensional Coulomb
potential $V(r) = - Ze^2/r$. The kinetic energy $K = \varepsilon - V(r)$ is
positive for $r < r_* \equiv Ze^2/|\varepsilon|$ and negative for $r > r_*$.
If the Dirac particle wavelength $\lambda = \hbar v_F /|\varepsilon|$ ($v_F$ 
being the Fermi velocity) is smaller than $r_*$, which occurs if 
$Z>\hbar v_F/e^2$, then the particle can be trapped inside $r_*$ but only for 
a finite amount of time before escaping due to the Klein-tunneling 
mechanism~\cite{beenakker:2008}. Since the ratio $r_*/\lambda$ is 
independent of the energy, an infinite number of such quasibound states are 
possible~\cite{SKL:2007b}. If one plots the local density of states (LDOS) 
versus the energy near the zero energy point, infinite oscillations can occur,
which is the defining characteristic of atomic collapse.

In this paper, we study particles confined on a curved space and uncover a 
class of quantum states similar to those that occur in atomic collapse. In 
general, the characteristics of quantum states on a curved surface constitute 
a fundamental problem in physics~\cite{OISYK:2012}. To derive the 
Schr\"{o}dinger equation governing the motion of a particle on a curved 
surface, an earlier approach was due to DeWitt~\cite{DeWitt:1957}, which was 
based on the quantization of the classical 2D Lagrangian. A difficulty with 
this method was that the particles are treated as intrinsically moving in the 
2D space, thereby generating the dilemma of ``operator ordering ambiguity'' 
that, for a classical function, multiple representative quantum operators may 
exist. The approach articulated by Jensen, Koppe, and da Costa 
(JKC)~\cite{JK:1971,daCosta:1981} overcomes this difficulty, where the
Schr\"{o}dinger equation was derived starting from the 3D Euclidean position
space followed by a reduction to 2D curved surface through an infinitesimally narrow confining potential locally normal to the surface. As a result, a general
feature of the Schr\"{o}dinger equation on a curved surface is a potential
term due to the intrinsic curvature of the 2D surface, and thus the so-called 
geometric potential. This approach has an experimental basis as the effects of 
the geometric potential on the quantum states have been observed experimentally
in electronic systems~\cite{NKKS:2001,OISYK:2012} and photonic topological
crystals~\cite{SDHKNTL:2010}. In fact, the JKC approach has become the
standard tool to study quantum mechanics on curved
surfaces~\cite{EM:2001,GW:2005,ZZW:2007,FC:2008,SYO:2009}.

To be concrete, we study a conic surface with its apex physically 
infinitesimally truncated in the sense that a circular region about the apex
with size of only one or two $\mathring{A}$ is removed, as shown in
Fig.~\ref{fig:cone_shape}. We employ the JKC method to derive the radial 
Schr\"{o}dinger equation on the truncated conic surface~\cite{FM:2008} and 
identify an effective potential that has an inverse squared dependence on 
the distance from the apex of the cone. This potential has a geometric origin,
which can be attractive or repulsive depending on the angular momentum quantum 
number. The analytical solutions of the Schr\"{o}dinger equation contain both 
bound and scattering states. Surprisingly, we uncover a class of abnormal 
scattering states that characteristically resemble the states underlying atomic
 collapse in a 2D system, e.g., in graphene. Since these unusual states are 
purely due to the curved geometry without the presence of any heavy nucleus, 
they are geometry-induced. Quantitatively, the ``collapse'' nature of these
states are established through the behavior of the LDOS, which we find exhibits
infinite oscillations - the defining characteristic of atomic collapse.  
Strictly speaking, they are only ``collapse-like'' states because atomic 
collapse is a relativistic quantum phenomenon but these states have a purely 
non-relativistic quantum origin. At the minimal risk of confusion, we still 
use the term ``collapse'' for convenience. To draw a stronger analogy of
these states with those in atomic collapse, we develop a qualitative analysis 
of the classical trajectories corresponding to the geometry-induced collapse 
states. Furthermore, we articulate possible experimental schemes to observe 
the exotic quantum states with a purely geometric origin. 
In terms of basic physics, our finding provides useful insights into the 
nature of quantum states in the curved space. With respect to applications,
our results suggest that wavefunction collapse should be an important factor 
of consideration in designing and developing nanodevices, because curved 
geometry has become increasingly common in nanoscience and nanotechnology.

Our main code is uploaded to GitHub: https://github.com/liliyequantum/Geometry-induced-wave-function-collapse.
\section{Schr\"{o}dinger equation on a truncated conic surface} \label{sec:model}

The starting point in studying the quantum dynamics of a particle on a 2D curved
surface is to derive the Schr\"{o}dinger equation on the surface. A previous
method was based on the idea of confining potential~\cite{JK:1971,daCosta:1981},
where one starts from the Schr\"{o}dinger equation in the 3D Euclidean space
and applies some appropriate potential to constrain the particle motion to
the curved surface. As a result, the Schr\"{o}dinger equation constrained on
a 2D curved surface defined by the metric tensor $g_{\mu\nu}$ can be written as~\cite{JK:1971,daCosta:1981}
\begin{equation} \label{eq:Schrodinger_1}
-\frac{\hbar^{2}}{2M}\left[\frac{1}{\sqrt{g}}\partial_{\mu}\left(\sqrt{g}g^{\mu\nu}
\partial_{\nu}\right)\right]\Psi+V_{G}\Psi=E\Psi,
\end{equation}
where $M$ is the particle mass, $g^{\mu\nu}$ is the contravariant component
of $g_{\mu\nu}$, $g=\mbox{det}\,g_{\mu\nu}$ and $V_{G}$ is a scalar geometric
potential given by
\begin{equation} \label{eq:Geo_Pot}
	V_{G}=-\frac{\hbar^{2}}{2M}(K_m^{2}-K),
\end{equation}
where $K_m$ and $K$ are the mean and Gaussian curvatures of points on the
surface, respectively, which characterize the internal and external geometric
properties of the surface. Note that $V_G$ has a pure geometric origin and it
is independent of any externally applied potential (if any). The quantum
properties of the normal mode $\chi_n$ in the perpendicular direction of the surface are governed by
\begin{equation} \label{eq:normal_mode}
-\frac{\hbar^{2}}{2M}\frac{\partial^{2}\chi_{n}}{\partial q_{n}^{2}}+V(q_{n})\chi_{n}=E_{n}\chi_{n},
\end{equation}
where $q_{n}$ is the coordinate normal to the surface and $V(q_{n})$ is the
confining potential that constrains the particle to the interface.

\begin{figure*} [ht!]
\centering
\includegraphics[width=\linewidth]{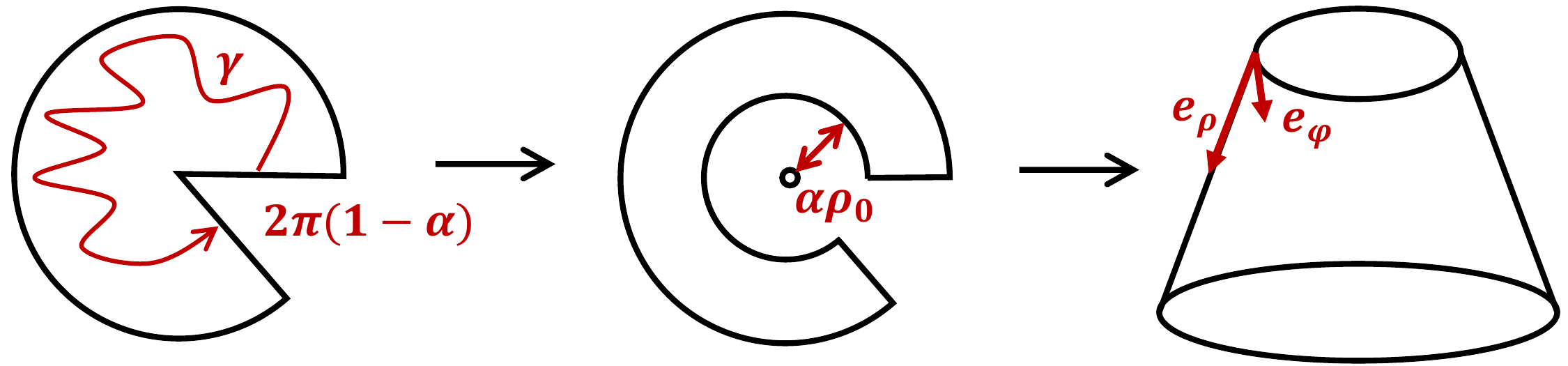}
\caption{A truncated conic surface with angular deficit $2\pi(1-\alpha)$ for 
$0 < \alpha < 1$. The truncation is physically infinitesimal in the sense
that the truncated distance away from the apex of the cone $\rho_0$ is chosen 
to have the size of only one or two atoms: $\rho_{0} \approx 2\mathring{A}$.}
\label{fig:cone_shape}
\end{figure*}

To be concrete, we consider the solution of the Schr\"{o}dinger equation
on a conic surface. A truncated cone can be obtained by a ``cut-and-glue''
process from a sheet of paper, as shown in Fig.~\ref{fig:cone_shape}. The
distance away from the apex of the cone in the circular cross section is 
denoted as $\rho\in[\rho_0,\infty)$, where the part of the cone with 
$\rho < \rho_0$ is removed. The truncation is physically infinitesimal in the 
sense that $\rho_0$ is chosen to be the size of one or two atoms, e.g.,
$\rho_{0} \approx 2\mathring{A}$. The line element or metric on a truncated 
cone is
\begin{equation} \label{eq:ds_cone}
ds^{2}=d\rho^{2}+\alpha^{2}\rho^{2}d\varphi^{2},
\end{equation}
where $\varphi\in[0,2\pi)$ and $2\pi\alpha$ ($0 < \alpha < 1$) is the sector 
angle of the corresponding solid angle of the cone. At $\rho = \rho_0$,
there is a hard wall boundary condition: $\psi|_{\rho_{0}}=0$, so
the wavefunction does not extend into the forbidden region $\rho < \rho_0$.
Since, as demonstrated in Appendix~\ref{appendix_A}, the geometric potential
induced by the mean and Gaussian curvatures has a singularity at $\rho = 0$,
the hard-wall boundary condition at $\rho = \rho_0$ removes this singularity
- a physically meaningful setting. 

The Schr\"{o}dinger Hamiltonian on a truncated conic surface 
becomes~\cite{FM:2008}
\begin{equation}\label{eq:2d equation}
H=-\frac{\hbar^{2}}{2M}\left[\frac{1}{\rho}\frac{\partial}{\partial\rho}\left(\rho\frac{\partial}{\partial\rho}\right)+\frac{1}{\alpha^{2}\rho^{2}}\frac{\partial^{2}}{\partial\varphi^{2}}\right]+V_G,
\end{equation}
where the geometry-induced potential is given by
\begin{equation} \label{eq:V_G_cone}
V_G=-\frac{\hbar^{2}}{2M}\left(\frac{1-\alpha^{2}}{4\alpha^{2}\rho^{2}}\right).
\end{equation}
Because of the circular symmetry of the conical geometric potential field, the
angular momentum $l$ is a good quantum number. The wavefunctions can thus be
naturally written in terms of the angular-momentum eigenstates $e^{il\varphi}$
as
\begin{equation}
\Psi(\textbf{r})=\psi(\rho)e^{il\varphi}
\end{equation}
with $l=0, \pm 1, \cdots$ ($l\in\mathbb{Z}$). In the angular momentum
representation, the Schr\"{o}dinger equation reduces to the following radial 
equation:
\begin{equation} \label{eq:RE}
\left[-\frac{\hbar^{2}}{2M}\frac{1}{\rho}\frac{d}{d\rho}\left(\rho\frac{d}{d\rho}\right) + U_G(\rho)\right]\psi(\rho)=E\psi(\rho),
\end{equation}
where
\begin{equation} \label{eq:UG}
U_G \equiv \frac{\hbar^{2}}{2M}\frac{\tilde{\nu}^{2}}{\rho^{2}}
\end{equation}
with
\begin{equation} \label{eq:tilde_nu}
\tilde{\nu}^{2}(\alpha,l) \equiv \frac{l^{2}}{\alpha^{2}}-\frac{1-\alpha^{2}}{4\alpha^{2}}.
\end{equation}
The first term in Eq.~\eqref{eq:tilde_nu} arises from the conical metric and
the second term originates from the mean curvature of the cone, a quantum
geometric potential.

Equations~\eqref{eq:UG} and \eqref{eq:tilde_nu} indicate that, for zero
angular momentum $l = 0$, the geometry-induced potential is attractive.
In this case, bound states will naturally arise~\cite{de:2005,FM:2008}. 
However, in spite of the attractive nature of the potential, a class of 
unusually extended or scattering states can emerge, and we will show below 
that they closely resemble the quantum states characteristic of atomic collapse
(the main result of this paper). For nonzero angular momentum states 
$|l| \ge 1$, the potential is repulsive, so the resulting scattering states 
are of the conventional type.

To simplify notations, we transfer the term $\sqrt{2ME}\rho/\hbar$ into the 
dimensionless form $\sqrt{\epsilon}r$ with the requirement 
$\sqrt{2ME_{0}}\rho_{0}/\hbar \approx 1$, where $\epsilon\equiv E/E_{0}$ and
$r\equiv\rho/\rho_{0}$. We assume $M$ to be the electron mass and consider 
$E_{0} = 1$ eV. The cutoff radial size is $\rho_{0}=1.93 \approx 2$ {\AA}.

\section{Characteristically distinct eigenstates} \label{sec:states}

Analytically solving Eqs.~\eqref{eq:RE}, \eqref{eq:UG} and \eqref{eq:tilde_nu},
we obtain three types of eigenstates: bound states and wavefunction collapse
states at zero angular momentum as well as conventional scattering states at 
finite angular momenta.

\subsection{Bound states} \label{subsec:bound_states}

\begin{figure} [ht!]
\centering
\includegraphics[width=\linewidth]{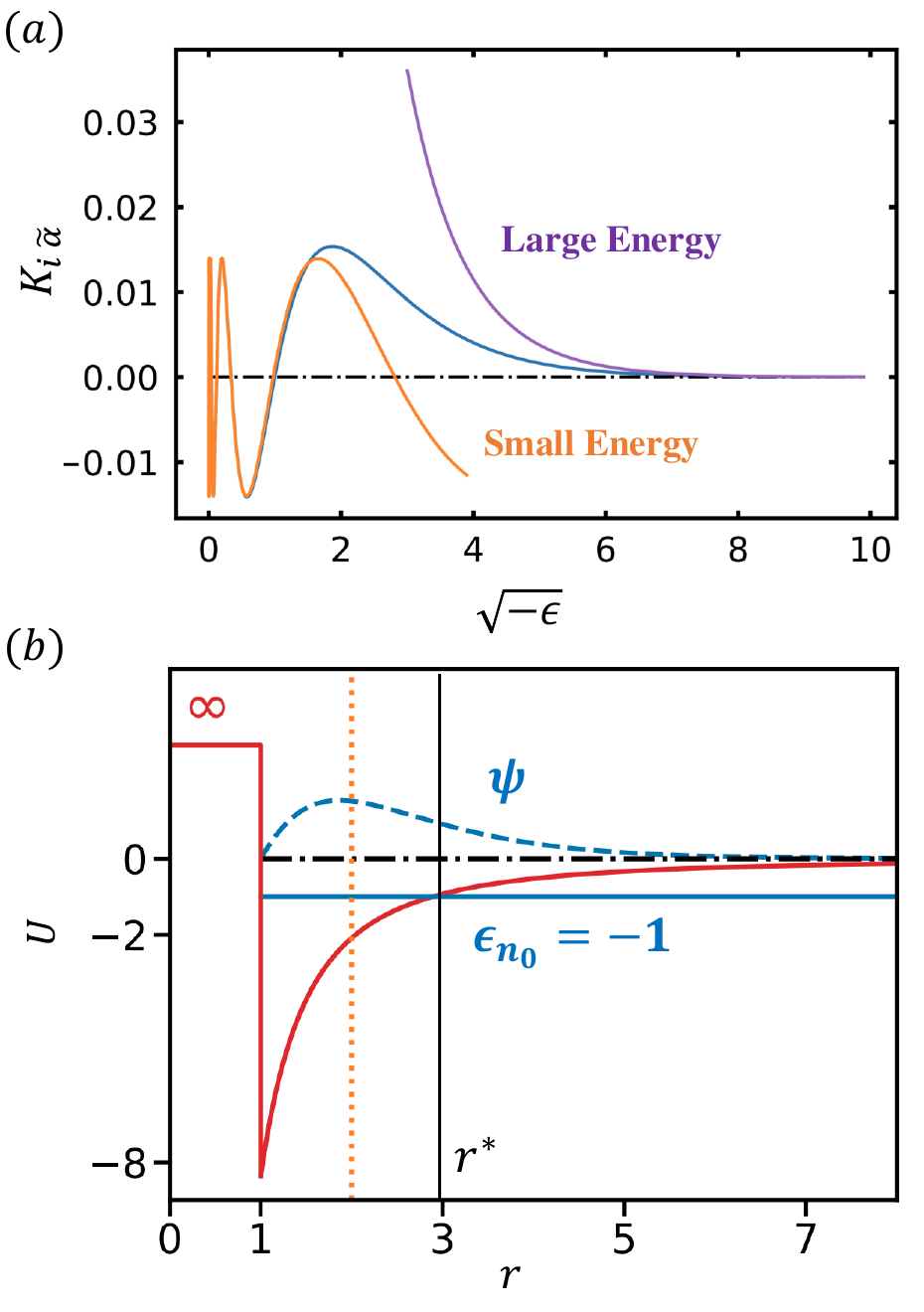}
\caption{Bound states for $\alpha=1/6$. (a) The bound states determined by the zeros of the function $K_{i\tilde{\alpha}}$(the middle curve), whose asymptotic
behaviors in the small and large energy regimes are given by
$\sin\left(\tilde{\alpha}\ln\left(x/2\right)-\phi_{\tilde{\alpha},0}\right)$
(the oscillatory curve for a small energy) and $e^{-x}/\sqrt{x}$ (the upper 
curve for a large energy), respectively, where $x=\sqrt{-\epsilon}$. (b) The 
ground state wavefunction (dashed curve) with the energy $\epsilon_{n_{0}}=-1$ 
(the solid horizontal line) for the potential of the form
$-\hbar^{2}\tilde{\alpha}^{2}/\left(2M\rho_{0}^{2}r^{2}\right)$ (solid trace) 
for $r \geq 1$. There is a hard wall at $r = 1$. The depth 
of the potential well - the minimal potential for the whole region, is about 
$U\approx-8$. The vertical dotted line denotes the ``center of mass'' 
$\langle r \rangle$ of the wavefunction, which is to the left of $r^{*}$, the 
classical forbidden region. All quantities plotted are dimensionless.}
\label{fig:bound_states}
\end{figure}

For $l=0$ and $E<0$, the quantum particle is effectively under the inverse 
square attractive potential and will be confined around the origin. Using the 
general solution of the Bessel equation of the imaginary order and the 
imaginary argument~\cite{dunster1990bessel}
\begin{equation}
   \psi\left(\textbf{r}\right)=AK_{i\nu}\left(x\right)+BL_{i\nu}\left(x\right),
\end{equation}
and considering the divergence of function $L_{i\nu}(x)$ at infinity, we have
that the solutions of Schr\"{o}dinger equation for $r\geq1$ and
$\alpha\in(0,\;1)$ are non-normalized bound states, which can be 
written as
\begin{equation}
\psi_{0,\epsilon_{n}}\left(\textbf{r}\right)=K_{i\tilde{\alpha}}\left(\sqrt{-\epsilon_{n}}r\right),
\end{equation}
where the new notation 
\begin{align} \nonumber
	\tilde{\alpha}\equiv\sqrt{1-\text{\ensuremath{\alpha^{2}}}}/\left(2\alpha\right)
\end{align}
is introduced to emphasize the imaginary order of the Bessel functions for 
zero angular momentum. 
Applying the boundary condition, we have that the zeros of $K_{\nu}(x)$
determine the discrete energy spectrum. In particular, at $\rho=\rho_{0}$ or 
$r=1$, applying the hard wall boundary condition leads to
\begin{equation}
K_{i\tilde{\alpha}}\left(\sqrt{-\epsilon_{n}}\right)=0.
\end{equation}
Figure~\ref{fig:bound_states} shows the zeros of the function
$K_{i\tilde{\alpha}}\left(\sqrt{-\epsilon}\right)$. For $\sqrt{-\epsilon}\rightarrow 0$, we have~\cite{dunster1990bessel}
\begin{equation}\label{eq:expan_k}
K_{i\tilde{\alpha}}\left(\sqrt{-\epsilon}\right)\rightarrow\sin\left(\tilde{\alpha}\ln\left(\sqrt{-\epsilon}/2\right)-\phi_{\tilde{\alpha},0}\right)=0,
\end{equation}
where $\phi_{\tilde{\alpha},0}=\arg\left\{\Gamma\left(1+i\tilde{\alpha}\right)\right\}$ and $\Gamma$ is the gamma function. The dimensionless eigenenergy 
spectrum is given by
\begin{equation} \label{eq:approx_bound_energer}
	\epsilon_{n}\approx-4\exp\left[2\left(-n\pi+\phi_{\tilde{\alpha},0}\right)/\tilde{\alpha}\right], 
\end{equation}
where for $\alpha\in[0.15,1)$, we have $n\in N^+$ and the ground state
corresponds to $n_{0}=1$, while for $\alpha\in(0,0.15)$, the approximation in
Eq.~\eqref{eq:expan_k} is invalid due to the increasing value of the 
ground-state energy. In this case, the minimal integer $n_{0}$ is less than 
one and the ground-state energy $\epsilon_{n_{0}}$ is smaller than the 
approximate value. For $\alpha\in(0,1)$, the whole eigenenergy spectrum 
$\epsilon_{n}$ goes from a finite negative value to $0^-$. Since 
$\sqrt{-\epsilon}=\sqrt{-2ME}\rho_{0}/\hbar$, the corresponding bound state 
energy spectrum becomes
\begin{equation}\label{eq:energy spectrum}
	E_{n}\approx\frac{\hbar^{2}\epsilon_{n}}{2M\rho_{0}^{2}},
\end{equation}
which is consistent with the result in Refs.~\cite{de:2005,FM:2008} with the 
approximation $\phi_{\tilde{\alpha},0}/\tilde{\alpha}\approx-\gamma$, where 
$\gamma$ is the Euler constant. There are then an infinite number of bound 
states.

For $\alpha=1/6$, the maximal zero root of 
$K_{i\tilde{\alpha}}(\sqrt{-\epsilon})$ occurs at $n=1$, which corresponds to 
the ground state. The difference in the energy level decreases 
as $n$ increases from $1$ to $\infty$. Since the function 
$K_{i\tilde{\alpha}}(\sqrt{-\epsilon})$ exhibits infinite oscillations near 
the zero energy point, as shown in Fig.~\ref{fig:bound_states}(a), there are 
an infinite number of bound states whose energy spectrum converges to zero 
$0^{-}$, near which the spectrum is quasi-continuous, corresponding to the
semiclassical regime. In the vicinity of the virtual zero root (corresponding
to $n=0$), the asymptotic behavior of $K_{i\tilde{\alpha}}(\sqrt{-\epsilon})$
is approximately exponential. For $r\rightarrow\infty$, we have
\begin{equation}
K_{i\tilde{\alpha}}\left(\sqrt{-\epsilon_{n}}r\right)\sim\sqrt{\frac{\pi}{2\sqrt{-\epsilon_{n}}r}}e^{-\sqrt{-\epsilon_{n}}r}.
\end{equation}
Figure~\ref{fig:bound_states}(b) shows, for $\alpha=1/6$, the wavefunction of
the ground state of energy $\epsilon_{n_{0}}=-1$. Using
Eq.~(\ref{eq:energy spectrum}) and considering that $\epsilon_{n}$ is
independent of $\rho_{0}$, we have $E_{n_{0}}\rightarrow-\infty$ for
$\rho_0\rightarrow 0$. In this case, the ground state corresponds to the
classical picture of the falling of the particle into the center as 
$\rho_0\rightarrow 0$ (an analogous situation was discussed by 
Landau~\cite{LL:book}). In principle, for $\rho_{0}\rightarrow 0$, all bound 
states with a finite energy correspond to classical trajectories falling to 
the center (to be analyzed in Sec.~\ref{sec:classical}).

\subsection{Scattering states with geometry-induced wavefunction collapse} \label{subsec:collapse_states}

For $l=0$ and $E>0$, a particle on the truncated conic surface experiences an 
equivalent inverse square attractive potential as for the case of bound states 
discussed in Sec.~\ref{subsec:bound_states}. In this case, the solutions are 
scattering states that exhibit infinite oscillations with energy near the zero 
energy point. In particular, using the general solution of the Bessel equation 
of the imaginary order and the real argument~\cite{dunster1990bessel}, we write
the real solution for $r\geq1$ and $\alpha\in(0,\;1)$ as
\begin{equation} \label{eq:collapse_states}
\psi_{0,\epsilon}\left(\textbf{r}\right)=AF_{i\tilde{\alpha}}\left(\text{\ensuremath{\sqrt{\epsilon}r}}
\right)-BG_{i\tilde{\alpha}}\left(\text{ \ensuremath{\sqrt{\epsilon}r}}\right),
\end{equation}
where the functions $F_{i\tilde{\alpha}}$ and $G_{i\tilde{\alpha}}$ are
linear combinations of the Hankel's functions of the first and second kind
[given by Eqs.~\eqref{eq:F} and \eqref{eq:G} in Appendix~\ref{appendix_B},
respectively], and the coefficients $0 \le A \le 1$ and $0 \le B \le 1$ are
\begin{align}\label{eq:AB_collapse}
	A & =\frac{G_{i\tilde{\alpha}}\left(\sqrt{\epsilon}\right)}{\sqrt{G_{i\tilde{\alpha}}^{2}\left(\sqrt{\epsilon}\right)+F_{i\tilde{\alpha}}^{2}\left(\sqrt{\epsilon}\right)}}, \nonumber\\
	B &=\frac{F_{i\tilde{\alpha}}\left(\sqrt{\epsilon}\right)}{\sqrt{G_{i\tilde{\alpha}}^{2}\left(\sqrt{\epsilon}\right)+F_{i\tilde{\alpha}}^{2}\left(\sqrt{\epsilon}\right)}}.
\end{align}

\begin{figure} [ht!]
\centering
\includegraphics[width=\linewidth]{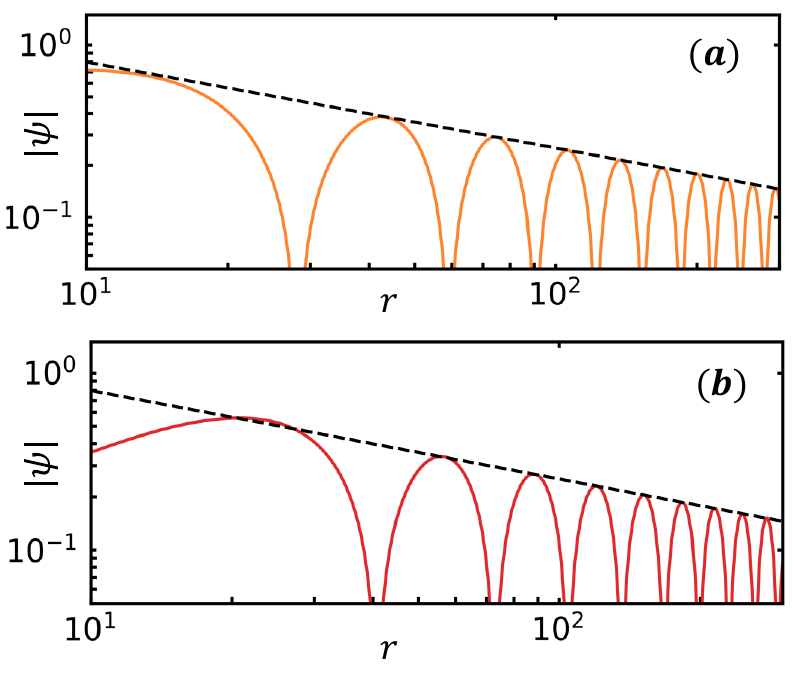}
\caption{Normalization of the quantum states at large distances from the 
conical apex: (a) a collapse state at zero angular momentum ($l=0$ and $E>0$) 
and (b) a conventional scattering state at a finite angular momentum ($|l|>0$ 
and $E>0$). The demensionless distance on the horizontal axis is defined at the end of section \ref{sec:model}.}
\label{fig:normalization}
\end{figure}

The solution in Eq.~\eqref{eq:collapse_states} satisfies the boundary condition
$\psi|_{r=1}=0$ and extends to infinity with proper normalization, as shown
in Fig.~\ref{fig:normalization}(a). For $r\rightarrow\infty$, the functions
$F_{i\tilde{\alpha}}\left(\sqrt{\epsilon}r\right)$ and
$G_{i\tilde{\alpha}}\left(\sqrt{\epsilon}r\right)$ tend to the conventional
Bessel's functions:
\begin{align}\label{eq:J_0}
	J_{0}\left(\sqrt{\epsilon}r\right)&\approx\sqrt{\frac{2}{\pi\sqrt{\epsilon}r}}\cos\left(\sqrt{\epsilon}r-\frac{\pi}{4}\right), \\ \label{eq:J_1}
J_{1}\left(\sqrt{\epsilon}r\right)&\approx\sqrt{\frac{2}{\pi\sqrt{\epsilon}r}}\sin\left(\sqrt{\epsilon}r-\frac{\pi}{4}\right).
\end{align}
respectively (Appendix B). Numerically, we find that for $\alpha\in[2/6,1)$, 
the asymptotic forms hold for $\sqrt{\epsilon}r>5$. For $\alpha<2/6$, the
asymptotic forms are valid for somewhat larger values of $\sqrt{\epsilon}r$.
For $\alpha\in(0,1)$ and fixed $r$, we have
the following asymptotic forms~\cite{dunster1990bessel} of
$F_{i\tilde{\alpha}}\left(\sqrt{\epsilon}r\right)$ and
$G_{i\tilde{\alpha}}\left(\sqrt{\epsilon}r\right)$ for $\epsilon\rightarrow0$:
\begin{align}
	F_{i\tilde{\alpha}}\left(\sqrt{\epsilon}r\right) &\sim\cos\left(\tilde{\alpha}\ln\left(\sqrt{\epsilon}r/2\right)-\phi_{\tilde{\alpha},0}\right), \\
	G_{i\tilde{\alpha}}\left(\sqrt{\epsilon}r\right)& \sim\sin\left(\tilde{\alpha}\ln\left(\sqrt{\epsilon}r/2\right)-\phi_{\tilde{\alpha},0}\right),
\end{align}
where $\phi_{\tilde{\alpha},0}=\arg{\left\{\Gamma(1+i\tilde{\alpha})\right\}}$ 
and $\Gamma$ is the gamma function. For $\epsilon\rightarrow 0$, the scattering
states given by Eq.~\eqref{eq:collapse_states} thus have the following form
\begin{equation} \label{eq:ZPIE}
\frac{A\sin\left(\tilde{\alpha}\ln\left(r\right)\right)}{\sqrt{B-C\cos^{2}\left
(\tilde{\alpha}\ln\left(\sqrt{\epsilon}/2\right)-\phi_{\tilde{\alpha},0}\right)}},
\end{equation}
where $r\geq 1$, $A=-\sqrt{2/\left(\tilde{\alpha}\pi\right)}$,
$B=\coth\left(\tilde{\alpha}\pi/2\right)$, and
$C=2/\sinh\left(\tilde{\alpha}\pi\right)$. In this near zero energy regime,
the wavefunction thus oscillates with the period $2\pi/\widetilde{\alpha}$ in 
a natural logarithmic scale. The resulting abnormal scattering states are 
effectively collapse states, corresponding to classically collapsing 
trajectories (see Sec.~\ref{sec:classical}).

\subsection{Scattering states with finite angular-momentum} \label{subsec:am_states}

For a nonzero angular momentum: $l \neq 0$, the overall inverse square
potential [Eq.~(\ref{eq:UG})] is repulsive, so the scattering states are 
conventional with a positive energy. The general solution of the Bessel 
equation of real order with a real argument is
\begin{equation}
    \psi\left(\textbf{r}\right)=AJ_{\nu}\left(x\right)+BY_{\nu}\left(x\right).
\end{equation}
so the scattering states for the whole energy region can be written as
\begin{equation} \label{eq:psi_normal}
\psi_{l,\epsilon}\left(\textbf{r}\right)=\left[AJ_{\tilde{\nu}}\left(\sqrt{\epsilon}r\right)-BY_{\tilde{\nu}}\left(\sqrt{\epsilon}r\right)\right]e^{il\varphi},
\end{equation}
for $l=\pm 1,\pm 2, \cdots$, where the coefficients $0 \le A \le 1$ and 
$0 \le B \le 1$ are given by
\begin{align}\label{eq:AB_conventional}
    A&=\frac{Y_{\tilde{\nu}}\left(\sqrt{\epsilon}\right)}{\sqrt{J_{\tilde{\nu}}^{2}\left(\sqrt{\epsilon}\right)+Y_{\tilde{\nu}}^{2}\left(\sqrt{\epsilon}\right)}},\nonumber\\
    B&=\frac{J_{\tilde{\nu}}\left(\sqrt{\epsilon}\right)}{\sqrt{J_{\tilde{\nu}}^{2}\left(\sqrt{\epsilon}\right)+Y_{\tilde{\nu}}^{2}\left(\sqrt{\epsilon}\right)}},
\end{align}
and the order of Bessel functions has the form 
\begin{equation}
\tilde{\nu}\left(\alpha,l\right)=\sqrt{\frac{l^{2}}{\alpha^{2}}-\frac{1-\alpha^{2}}{4\alpha^{2}}}.
\end{equation}
with $\alpha\in(0,1)$. Equation~\eqref{eq:psi_normal} is the exact 
analytical solution, where $Y_{\tilde{\nu}}(\sqrt{\epsilon}r)$ diverges at 
the boundary $r=1$ for $\epsilon\approx0$. In numerical simulations, we set 
the maximum cutoff as $Y_{\tilde{\nu}}\leq 100$, guaranteeing the hard-wall 
boundary condition at $r=1$. The maximal error of LDOS is of the order of 
$10^{-32}$ near the zero energy point and in the finite energy region 
$(0,10]$. Asymptotically, as shown in Fig.~\ref{fig:normalization}(b), 
the conventional scattering states can be normalized at infinity through 
the standard form
\begin{align} \label{eq:J_tilde}
J_{\tilde{\nu}}&\left(\sqrt{\epsilon}r\right)\sim\sqrt{\frac{2}{\pi\sqrt{\epsilon}r}}\cos\left(\sqrt{\epsilon}r-\frac{\tilde{\nu}\pi}{2}-\frac{\pi}{4}\right),\nonumber\\
	Y_{\tilde{\nu}}&\left(\sqrt{\epsilon}r\right)\sim\sqrt{\frac{2}{\pi\sqrt{\epsilon}r}}\sin\left(\sqrt{\epsilon}r-\frac{\tilde{\nu}\pi}{2}-\frac{\pi}{4}\right). 
\end{align}
We discuss two extreme cases among the three kinds of quantum states:
$\alpha\rightarrow 1$ and $\alpha\rightarrow 0$. For $\alpha\rightarrow1$ with
fixed $\rho_{0}$, the conic surface becomes a 2D plane with a hole of radius
$\rho_{0}$ at the center. The geometric potential vanishes because 
$\tilde{\alpha}^2=(1-\alpha^2)/(4\alpha^2)=0$. In this case, the bound states
disappear due to the zero depth of the potential well in the form of 
$\hbar^2\tilde{\alpha}^2/(2M\rho^2_{0}r^2)$, which is defined by the minimum
of the effective potential. The geometry-induced collapse states and scattering
states with finite angular momenta degenerate into the normal scattering 
states in the plane, which can be expressed as a linear combination of 
$J_{l}(\sqrt{\epsilon}r)$ and $Y_{l}(\sqrt{\epsilon}r)$ multiplied by  
$e^{il\varphi}$ for $l=0,\pm1,\pm2,...$ with 
\begin{align} \nonumber
	&\lim_{\tilde{\alpha}\rightarrow0} F_{i\tilde{\alpha}} \rightarrow J_{0}(\sqrt{\epsilon}r), \\ \nonumber
	&\lim_{\tilde{\alpha}\rightarrow0} G_{i\tilde{\alpha}}(\sqrt{\epsilon}r) \rightarrow Y_{0}(\sqrt{\epsilon}r), 
\end{align}
which have been verified numerically and are consistent with, e.g., Eq.~(3.3)
in Ref.~\cite{dunster1990bessel} and $\tilde{\nu}\rightarrow l$. For 
$\alpha\rightarrow 0$ with fixed $\rho_{0}$, the conic and cylindrical surface 
has an infinitesimally small radius. In this case, since 
$\tilde{\alpha}\rightarrow\infty$, the geometric potential is homogeneously 
infinite for the whole surface region. Because the potential is infinitely 
negative for zero angular momentum and infinitely positive for nonzero 
angular momenta, the wavefunctions simply vanish.

\section{Local density of states and demonstration of collapse states}

In general, the characteristics of the wavefunction depend on the distance from
the apex of the cone $r$ and the sector angle of a truncated cone as measured 
by $2\pi\alpha$, which can be studied through the LDOS. The general definition 
of LDOS~\cite{pereira2007coulomb} is
\begin{equation}
N\left(\epsilon,\textbf{r}\right) = \sum_{\epsilon'}\left|\Psi_{\epsilon'}\left(\textbf{r}\right)\right|^{2} \delta\left(\epsilon-\epsilon'\right) = \sum_{l=-\infty}^{+\infty}n_{l}\left(\epsilon,\textbf{r}\right),
\end{equation}
where $n_{l}\left(\epsilon,\textbf{r}\right)= \left|\psi_{l,\epsilon}\left(\textbf{r}\right)\right|^{2}$, a quantity that involves only the positive energy states.

\begin{figure} [ht!]
\centering
\includegraphics[width=\linewidth]{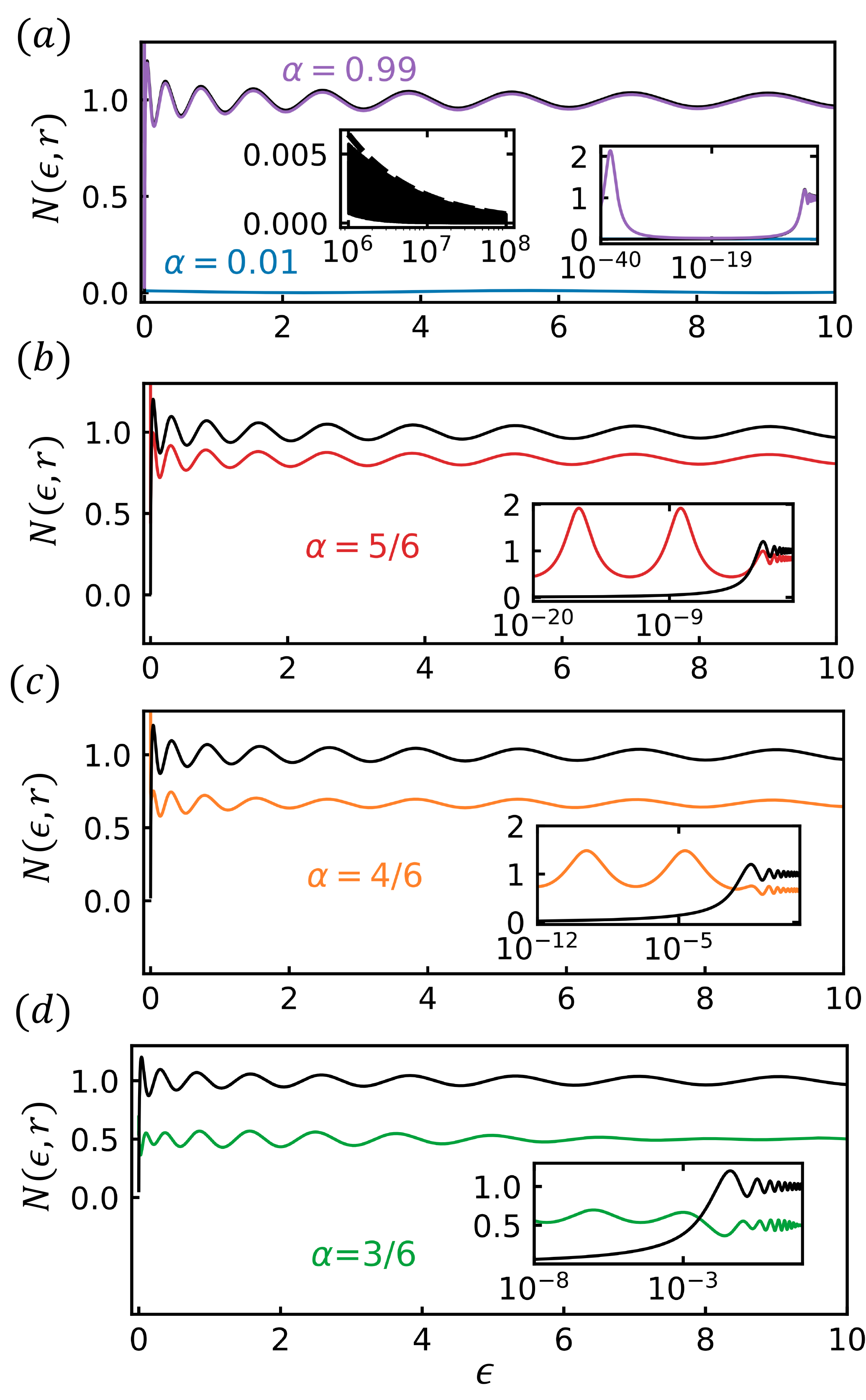}
\caption{Behavior of the total LDOS for different conic surfaces at $r=10$.
(a) LDOS for the two extreme cases: $\alpha = 0.99$ and $\alpha = 0.01$(the solid curves and the right inset for small energy),
where in the former the LDOS oscillate about the value one coincided with normal scattering states in the 2D plane with the hard 
hole $r=1$ when $\alpha=1$ and, in the latter, no quantum 
states exist to contribute to the LDOS. In a higher energy
region, the asymptotic behavior of the LDOS for $\alpha= 1$ is shown 
in the left inset. (b-d) LDOS plots for $\alpha = 5/6$, $4/6$ and $3/6$, 
respectively, where the top curve in each panel is for $\alpha = 1$. 
In these cases, the values around which the LDOS 
oscillates are between zero and one. In a small energy interval near zero, 
the LDOS exhibits infinite oscillations with the energy, as shown in the 
respective insets of the three cases. As argued in the text, in a near zero 
energy interval, the main contribution to the LDOS comes from the zero angular 
momentum states, where the infinite oscillations are indicative of the collapse
nature of these states. In each panel, the dimensionless energy $\epsilon$ is defined at the end of the section \ref{sec:model}.}
\label{fig:total_LDOS}
\end{figure}

\begin{figure} [ht!]
\centering
\includegraphics[width=\linewidth]{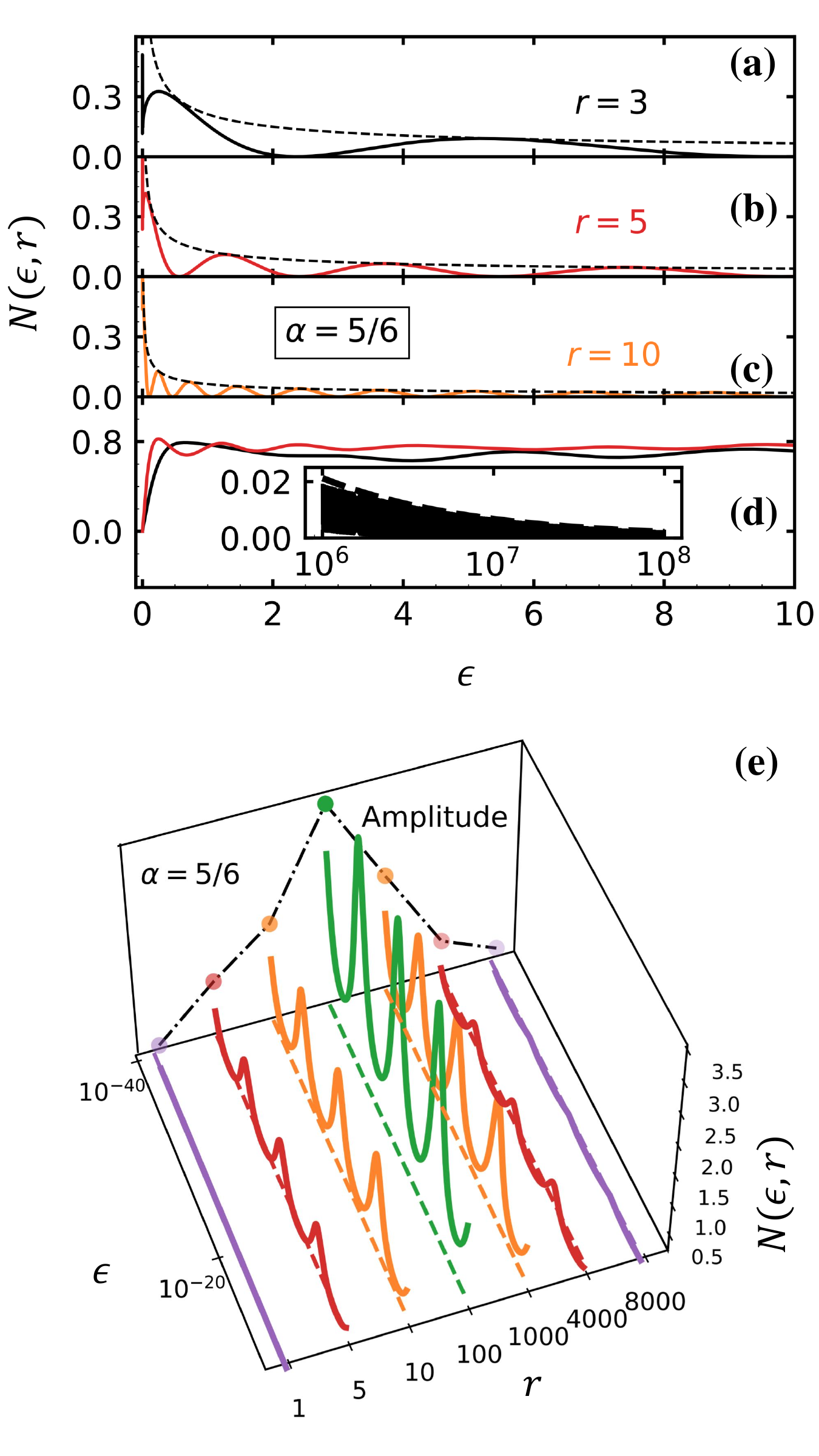}
\caption{Contributions to the LDOS from the collapse and conventional 
scattering states. (a-c) The contribution from the collapse states for three
distance values with the asymptotic behavior (dashed 
curves) in a large energy interval. (d) The contribution from 
the conventional scattering states at two distances $r=3,5$. The inset is for the decay behavior in the higher energy interval when $r=3$(solid cure) with the asymptotic dashed line.
(e) Oscillations of the LDOS due to the collapse states in a small near-zero 
energy interval for different values of the distance, as represented by a 3D 
plot of the LDOS in terms of both the energy and distance. The amplitudes of 
LDOS oscillations for different distances are projected to the 2D plane. All
quantities plotted are dimensionless.}
\label{fig:distance}
\end{figure}

\begin{figure*} [ht!]
\centering
\includegraphics[width=\linewidth]{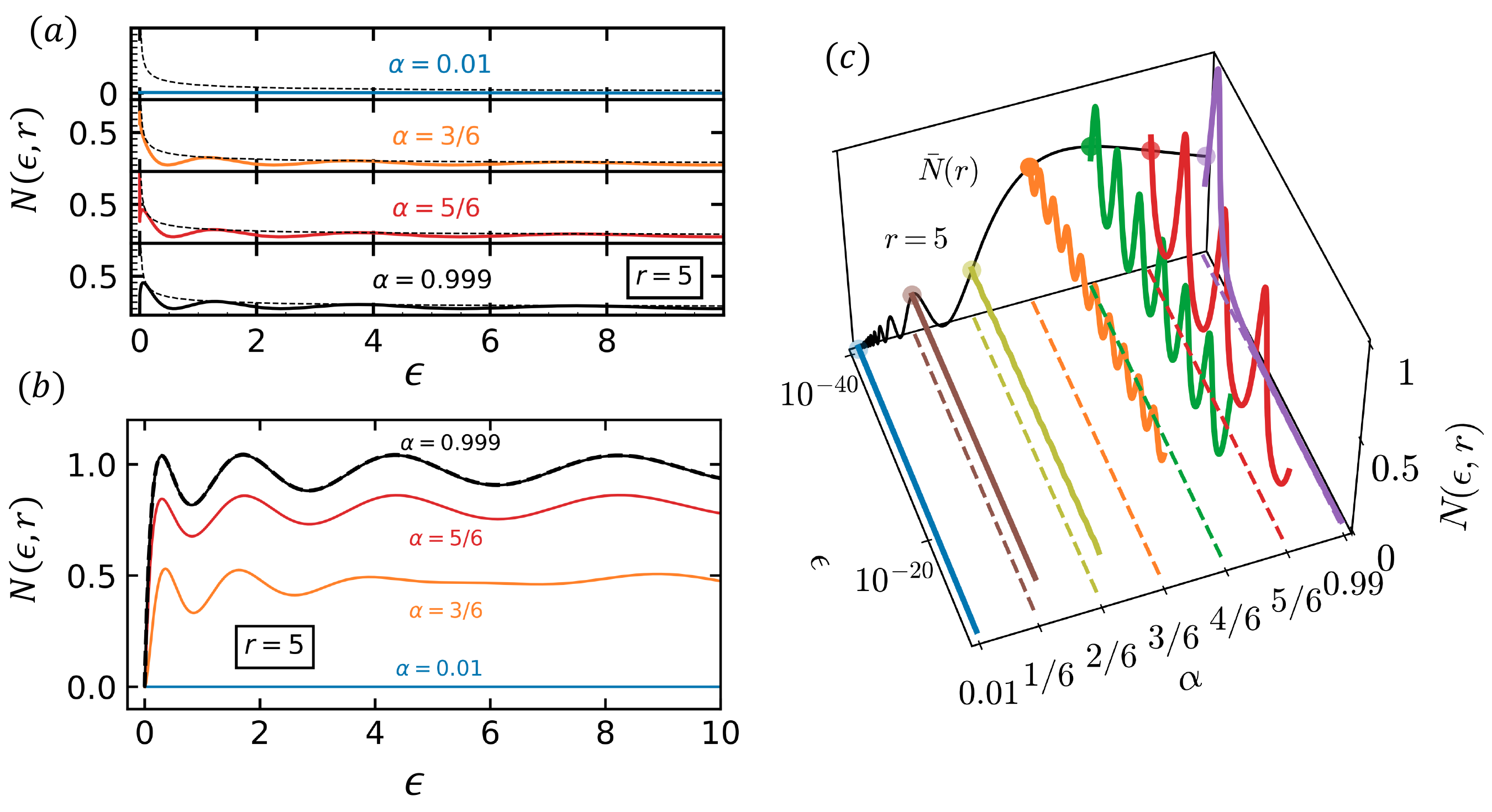}
\caption{Effect of different conic geometry on the LDOS. 
(a,b) LDOS versus energy contributed to by the collapse and conventional 
scattering states, respectively, for $r = 5$ and four different values of the 
sector angle $2\pi\alpha$. The dashed curves in (a) show the asymptotic 
behaviors in the high energy region. The dashed curve in (b) show the LDOS of 
the conventional scattering states in 2D plane with a hard hole. (c) The LDOS 
contributed to by the collapse states in an infinitesimal energy interval near 
zero, the frequencies of which decrease while the amplitudes increase as 
$\alpha$ increase from zero to one. The average LDOS $\bar{N}(r)$ over the near
zero energy point is plotted versus $\alpha$, which is projected to the 2D 
plane. The solid curve in the back plane is the prediction of 
Eq.~\eqref{eq:LDOS_bar} and the data points are the corresponding numerical 
results. All quantities plotted are dimensionless.}
\label{fig:alpha}
\end{figure*}

Evidence of the emergence of the collapse states is presented in
Figs.~\ref{fig:total_LDOS}(b-d) for $\alpha=5/6$, $4/6$, and $3/6$,
respectively, where the infinite oscillations of the LDOS are shown in the
corresponding insets. Note that, for $\alpha=3/6$, LDOS oscillations can be
seen in a relatively large energy region: $E\sim\mu eV$ (corresponding roughly
to the accessible resolution in the current experimental 
technology~\cite{schwenk:2020}). The results indicate that zero energy is the 
accumulation point of infinitely many resonances, a characteristic of the 
atomic collapse states~\cite{PNN:2007,SKL:2007a,SKL:2007b}. The collapse states
arise from the  conic surface for $\alpha$ values close neither to one nor to 
zero, as the energy interval in which the LDOS oscillations are pronounced
shrinks to zero for $\alpha\rightarrow 1$ and the LDOS is zero for 
$\alpha\rightarrow 0$.

For large energy $\epsilon\rightarrow\infty$, according to 
Eqs.~\eqref{eq:psi_normal}, \eqref{eq:AB_conventional} and \eqref{eq:J_tilde},
the norm square of the conventional scattering states with fixed angular 
momenta can be written as
\begin{align}\label{eq:conventional_asymp}
    \lim_{\epsilon\rightarrow\infty}|\psi_{l,\epsilon}|^{2}\rightarrow\frac{2}{\pi\sqrt{\epsilon}r}\sin^{2}\left(\sqrt{\epsilon}\left(1-r\right)\right),
\end{align}
which is independent of the angular momentum quantum number $l$ because of
the asymptotic relations:
\begin{align} \nonumber
        A &\rightarrow\sin\left(\sqrt{\epsilon}-\frac{\tilde{\nu}\pi}{2}-\frac{\pi}{4}\right) \ \ \mbox{and} \\ \nonumber
        B&\rightarrow\cos\left(\sqrt{\epsilon}-\frac{\tilde{\nu}\pi}{2}-\frac{\pi}{4}\right) 
\end{align}
with $\epsilon\rightarrow\infty$. The asymptotic LDOS of the conventional 
states for large energies is thus proportional to $N_l$: 
LDOS $\propto N_l\lim_{\epsilon\rightarrow\infty}|\psi_{l,\epsilon}|^{2}$. 
For $l\in[-50,0)\cup(0,50]$, we have $N_l=100$ for conventional scattering 
states, as shown in the inset of Fig.~\ref{fig:distance}(d). For 
$l\in[-50,50]$, we have $N_l=101$ to include the degenerate collapse 
states for $\alpha$ close to one, as shown in the inset of 
Fig.~\ref{fig:total_LDOS}(a). For the collapse states with $\alpha\in(0,1)$, 
the asymptotic LDOS has the form $2/(\pi\sqrt{\epsilon}r)$, 
as shown in Figs.~\ref{fig:distance}(a-c) and \ref{fig:alpha}(a) based on
Eqs.~\eqref{eq:AB_collapse}, \eqref{eq:J_0} and \eqref{eq:J_1}.

As $\alpha$ decreases from one to zero, the value around which the total 
LDOS oscillates reduces from one to zero, as shown in 
Figs.~\ref{fig:total_LDOS}(a-d). The main reduction comes from the 
conventional scattering states shown in Figs.~\ref{fig:alpha}(a,b). This 
can be argued heuristically, as follows. For a fixed distance from the 
conical apex, the wavefunctions 
$F_{i\tilde{\alpha}}\left(\text{\ensuremath{\sqrt{\epsilon}r}}\right)$ and 
$G_{i\tilde{\alpha}}\left(\text{\ensuremath{\sqrt{\epsilon}r}}\right)$ for 
the sufficiently large energy, e.g., $\sqrt{\epsilon}r>5$, tend to $J_{0}$ 
and $J_{1}$, respectively, regardless of the values of $\alpha$. Consequently,
the reduction does not occur for the collapse states, as shown in 
Fig.~\ref{fig:alpha}(a). For the conventional scattering states, given a 
finite energy interval, the high angular momentum states 
$J_{\tilde{\nu}}(\sqrt{\epsilon}r)$ will be pushed out of this energy 
interval into a higher energy region, leaving behind the low angular-momentum 
states to contribute to the total LDOS, as shown in Fig.~\ref{fig:alpha}(b). 
As a result, the value around which the LDOS oscillates will reduce with 
$\alpha$. In the extreme case of $\alpha$ decreasing to zero, the number of 
contributing states becomes zero.

To obtain a more comprehensive picture of the contribution of the collapse
states to the LDOS, we decompose it into two parts:
\begin{equation}
N\left(\epsilon,\textbf{r}\right)=\sum_{l=0}\bar{n}_{l}\left(\epsilon,\textbf{r}\right)+\sum_{l\neq0}n_{l}\left(\epsilon,\textbf{r}\right),
\end{equation}
where the first and second terms are the contributions from the collapse states
and the conventional scattering states, respectively. As an example, we fix
$\alpha = 5/6$ and examine the two types of contribution at different
distances from the apex of the cone. Figs~\ref{fig:distance}(a-c) show
the contribution to the LDOS from the collapse states for three distance
values, respectively, while Fig.~\ref{fig:distance}(d) displays the
contribution from the conventional scattering state at two distances.
The oscillations of the collapse-state contributed LDOS near the zero energy
point with the distance exhibit a different behavior, as shown in a 3D plot of
$N(\epsilon,r)$ versus the energy and distance, as exemplified in
Fig.~\ref{fig:distance}(e). In this interval of infinitesimal energies, the
oscillation amplitude of $N(\epsilon,r)$ depends on the distance $r$ in the form
of $\sin(\tilde{\alpha}\ln{r})$ as in Eq.~\eqref{eq:ZPIE}, described by the 2D projection in the 3D plot of Fig.~\ref{fig:distance}(e). The oscillation
frequency depends on $\alpha$ as determined by
\begin{displaymath}
\cos^{2}\left(\tilde{\alpha}\ln\sqrt{\epsilon}+C\left(\tilde{\alpha}\right)\right),
\end{displaymath}
providing an explanation of the observed same number of periods of oscillation
at different distances for the same energy range, as shown in Fig.~\ref{fig:distance}(e).

What is the effect of varying the sector angle $2\pi\alpha$ of the truncated 
cone on the LDOS? Figures~\ref{fig:alpha}(a,b) show, for fixed $r = 5$ and 
several values of $\alpha$, the LDOS versus the energy for the contributions 
from the collapse and conventional scattering states, respectively. In both 
cases, the number of oscillation periods is independent of the value of 
$\alpha$, as can be seen from Eqs.~\eqref{eq:J_0}, \eqref{eq:J_1} and 
\eqref{eq:conventional_asymp}. In an infinitesimal energy interval near zero,
Eq.~\eqref{eq:ZPIE} stipulates that the oscillation amplitude of the LDOS
associated with the collapse states enhances with $\alpha$ but the
oscillation frequency reduces, as shown in Fig.~\ref{fig:alpha}(c). For a near 
zero $\alpha$ value, e.g., $\alpha=0.01$, the oscillation amplitude is 
approximately zero. In the opposite extreme case, e.g., $\alpha=0.99$, the LDOS
exhibits a single oscillation and then approaches zero.

To be concrete, we define the average LDOS with respect to energy values near
the zero energy point $\epsilon\approx 0$. In this energy interval, the LDOS 
is mainly contributed to by the geometry-induced collapse states, which 
exhibits regular oscillations as stipulated by Eq.~\eqref{eq:ZPIE}. The 
average LDOS is given by 
\begin{align}
	\bar{N}(r) = \left[\max(N(\epsilon,r))+\min(N(\epsilon,r))\right]/2.
\end{align}
The 2D projection of the 3D plot in Fig.~\ref{fig:alpha}(c) shows the average 
value of the LDOS, where the black curve represents the theoretical formula 
obtained from Eq.~\eqref{eq:ZPIE}:
\begin{equation} \label{eq:LDOS_bar}
	\bar{N}(r) = \frac{1}{2}\left[\frac{1}{B}+\frac{1}{B-C}\right]A^{2}\sin^{2}\left[\tilde{\alpha}\ln\left(r\right)\right],
\end{equation}
where $A$, $B$ and $C$ are defined by Eq.~\eqref{eq:ZPIE}, which only depend 
on $\tilde{\alpha}$. The data points in Fig.~\ref{fig:alpha}(c) are from 
numerical simulations, which falls precisely on the theoretical curve.

\section{A classical picture} \label{sec:classical}

\begin{figure*} [ht!]
\centering
\includegraphics[width=\linewidth]{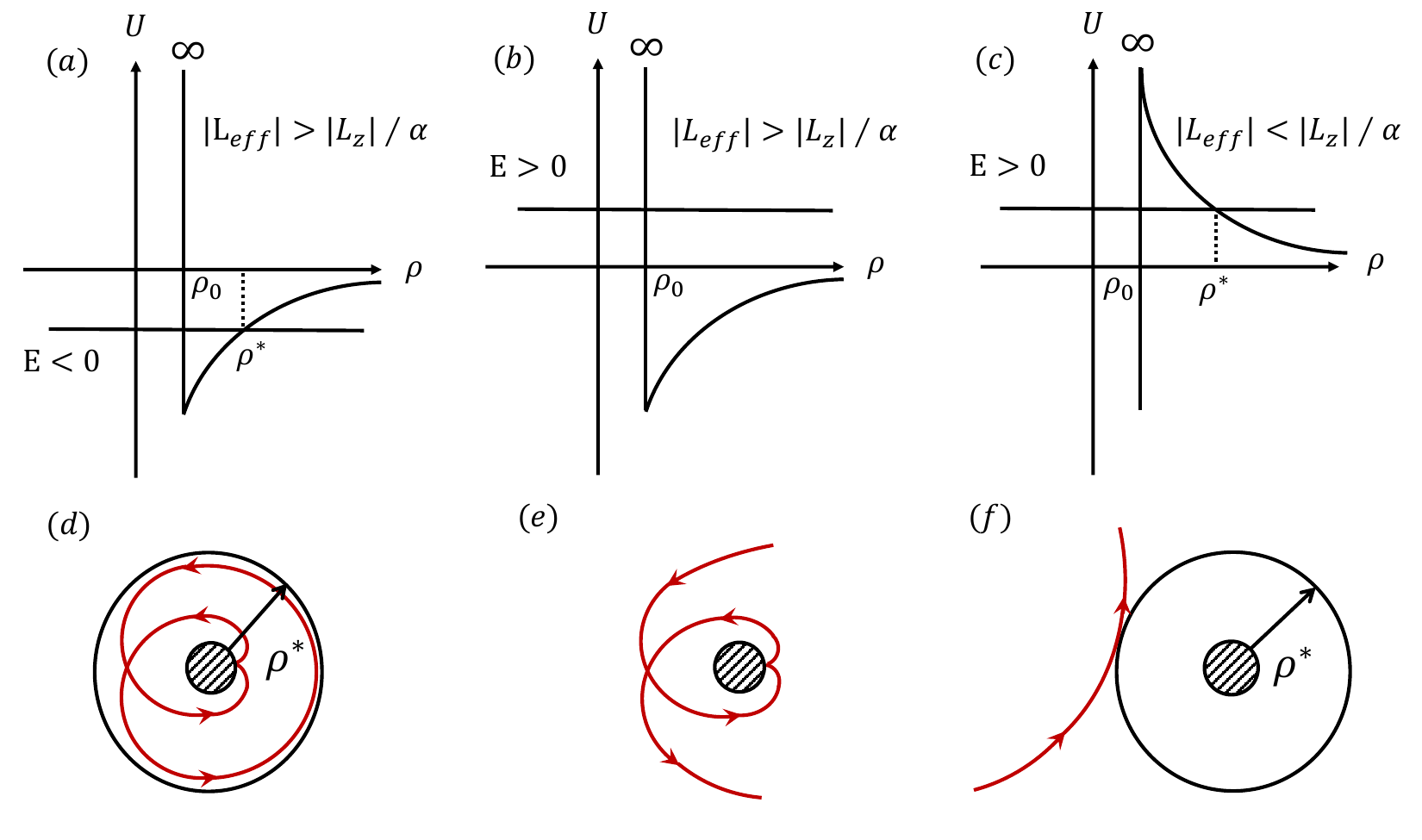}
\caption{Representative classical particle trajectories on a truncated conic 
surface from the Hamiltonian (\ref{eq:classical Hamiltonian0}).
(a-c) Three different potential profiles, and (d-f) the corresponding
classical trajectories. (a,d) For $|L_{eff}|>|L_{z}|/\alpha$ and 
$E< 0$, the effective potential is attractive and the particle is confined in 
the region $(\rho,\rho^*)$, corresponding to bound states. (b,e) For 
$|L_{eff}|>|L_{z}|/\alpha$ and $E>0$, the classical particle can 
collapse to $\rho_0$ in finite time but would eventually escape to infinity 
due to the reflective boundary condition at $\rho_0$. Geometry-induced 
wavefunction collapse occurs in this case. (c,f) For 
$|L_{eff|}<|L_{z}|/\alpha$ and $E > 0$, the effective potential is 
repulsive and the particle is confined in the region $(\rho^*,\infty)$, 
corresponding to the conventional quantum scattering case.}
\label{fig:CT}
\end{figure*}

To gain deep physical insights into the geometry-induced collapsed states, we
construct the corresponding classical picture, following the pioneering work
on geometric potential~\cite{shortley:1931,shytov2007atomic}. Consider a 
classical particle of mass $M$ moving on a 2D truncated conic surface, subject 
to an external potential. The effective classical potential is
$-L_{eff}^{2}/(2M\rho^{2})$, where $L_{eff}$ is the coefficient.
The classical Hamiltonian is 
\begin{equation}\label{eq:classical Hamiltonian0}
   H=\frac{\left|\textbf{p}\right|^{2}}{2M}- \frac{L_{eff}^{2}}{2M\rho^{2}}.
\end{equation}
The particle is constrained to move in the region $\rho>\rho_{0}$ with a hard 
wall at $\rho=\rho_{0}$. The classical linear momentum can be decomposed into
two parts: the radial and angular components, as
\begin{equation}
\left|\textbf{p}\right|^{2}=p_{\rho}^{2}+L_{z}^{2}/\left(\alpha^{2}\rho^{2}\right),
\end{equation}
where $L_{z}=\alpha\rho p_{\varphi}$ is the classical angular momentum 
characterizing the particle motion around the $z$ axis. Since the potential 
$-L_{eff}^{2}/(2M\rho^{2})$ results in a central force field, the 
angular momentum $L_{z}$ is conserved. The classical Hamiltonian can be 
expanded as
\begin{equation}\label{eq:classical Hamiltonian}
H=\frac{p_{\rho}^{2}}{2M}+\frac{1}{2M\rho^{2}} \left(\frac{L_{z}^{2}}{\alpha^{2}}-L_{eff}^{2}\right),
\end{equation}
where the quantities in the bracket of the second term are constants, so this 
term is effectively a potential function of $\rho$. Depending 
on $L_z$ and $L_{eff}$, this effective potential can be either 
positive or negative. The radial motion of the particle is thus completely
governed by the Hamiltonian \eqref{eq:classical Hamiltonian}.

For $|L_{eff}|>|L_{z}|/\alpha$, the second term in 
Eq.~(\ref{eq:classical Hamiltonian}) is negative and the attractively effective potential. For total negative energy 
$E<0$, the particle is trapped inside the region with radius 
$\rho \in (\rho_0,\rho^{*})$, where $p_{\rho}(\rho^{*})=0$, as shown in 
Fig.~\ref{fig:CT}(a). The particle spirals inward and 
reflects from the hard wall boundary at $\rho_0$, spirals outward, is 
pulled back by the attractive potential, begins to spiral inward again, 
and so on, as depicted in Fig.~\ref{fig:CT}(d). This type
of motion corresponds to the bound states in the quantum regime.

For $E>0$, in the $\rho \rightarrow \infty$ limit, there is a radial kinetic 
energy of the form $p_{\rho}^{2}/(2M)$. In this case, $\rho^{*}$ extends to 
infinity. As illustrated in Figs.~\ref{fig:CT}(b) and \ref{fig:CT}(e), the 
particle spirals inward toward the center from infinity, is reflected at the 
boundary $\rho_0$ and then spirals outward back to infinity. This is 
the classical picture of the geometry-induced collapse state with an infinitely
oscillating local density of states. 

For $|L_{eff}|<|L_{z}|/\alpha$, the second term in 
Eq.~(\ref{eq:classical Hamiltonian}) is positive and the repulsively
effective potential.
The motion of the particle is constrained in the region 
$\rho \in (\rho^*,\infty)$. As illustrated in Figs.~\ref{fig:CT}(c) and 
\ref{fig:CT}(f), it is not a falling trajectory and the particle is 
scattered away from $\rho^*$, corresponding to the conventional quantum 
scattering states. Note that $\rho_{0}$ is assumed to be 
sufficiently small so that the potential at $\rho=\rho_{0}$ can be regarded
as infinite. 
Furthermore, if we quantify $L_{z}$ as $l\hbar$ following the standard
procedure of quantization and assume $L_{eff}$ is equivalent to 
$\hbar\tilde{\alpha}$ in the quantum-classical correspondence, the effective 
potential in Eq.~(\ref{eq:classical Hamiltonian}) will have the same 
mathematical form as Eqs.~(\ref{eq:UG}) and (\ref{eq:tilde_nu}).

\section{Experimental feasibility of observing the geometry-induced collapse states}

\begin{figure} [ht!]
\centering
\includegraphics[width=\linewidth]{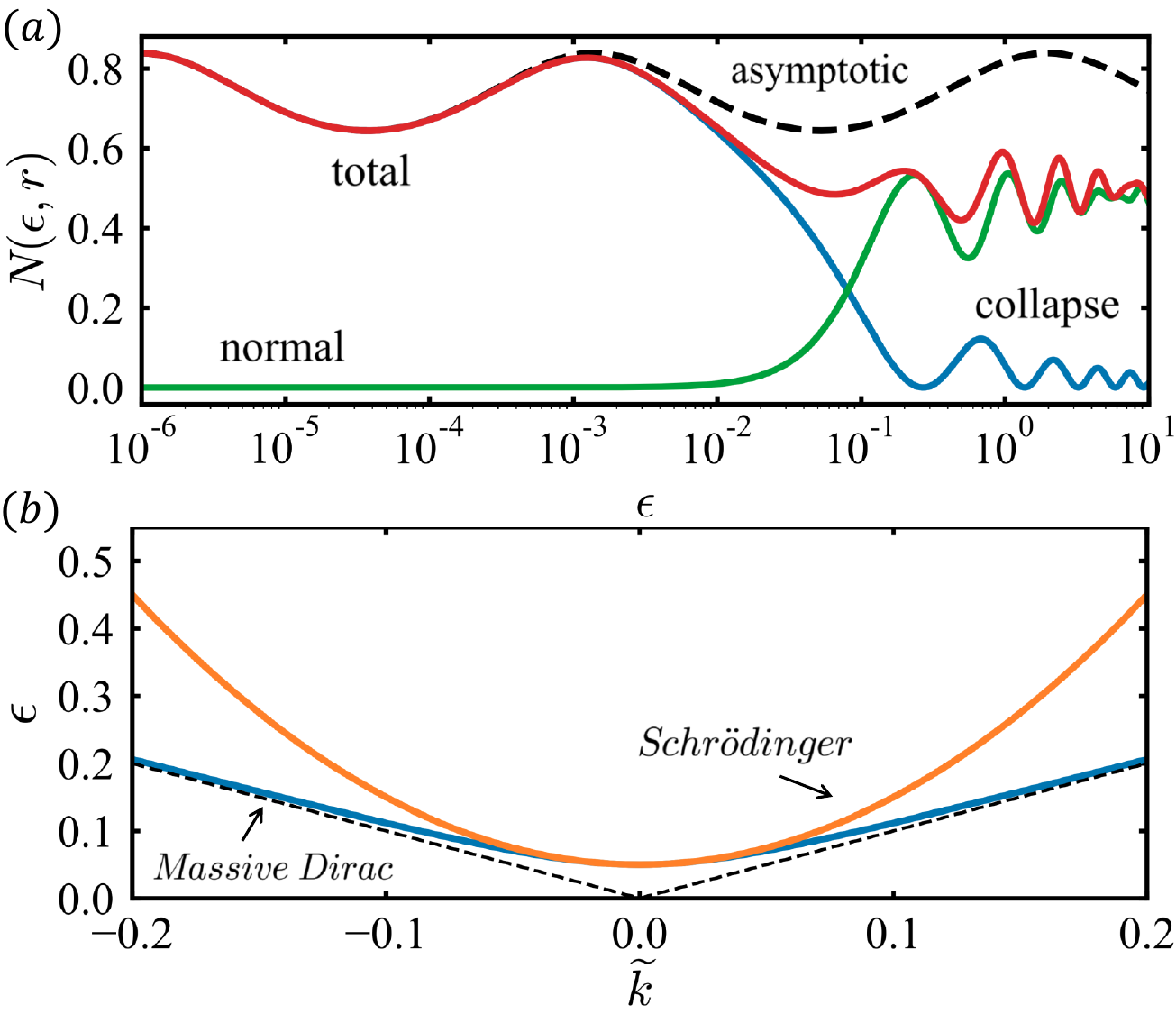}
\caption{Feasibility of experimentally observing 
geometry-induced wavefunction collapse through LDOS oscillations. (a) For 
the sector angle of the truncated cone $2\pi\alpha=\pi$, LDOS oscillations 
associated with total, collapse and normal states with the corresponding 
asymptotic behavior near zero energy, where $\epsilon\in[10^{-6},10^{-3}]$ 
corresponds to $\Xi\in2\times[1,10^3]$  $\mu$eV. These oscillations are 
experimentally feasible~\cite{KDTHLE:1997,terrones:2001} (see text for an 
analysis). (b) The energy-momentum dispersion relation of a massive Dirac 
Fermion and a Schr\"{o}dinger particle with the gap $\widetilde{\Delta}=0.05$,
corresponding to $\Delta=0.1$ eV.}
\label{fig:graphene_experimental_scheme}
\end{figure}

We analyze in detail the feasibility of observing the phenomenon of 
geometry-induced wavefunction collapse. A basic issue is
to measure the LDOS oscillations associated with the collapse phenomenon.  
With the development of the STM technology~\cite{BR:1987,HT:1987}, the 
LDOS can be detected by STM with tunneling current proportional to the 
LDOS of the surface at the position of the 
tip~\cite{tersoff:1985,li:1997,Wangetal:2013}. For example, a recent 
experimental work~\cite{schwenk:2020} reported the achievement of a 
$\mu eV$ tunneling resolution with in-operando measurement capabilities of STM,
making it feasible to observe the oscillations in the LDOS associated with the 
collapse state, as shown in Fig.~\ref{fig:graphene_experimental_scheme}(a), 
where the dimensional energy is about $\epsilon$ eV (defined below).

Since the experimental observation of atomic collapse was primarily achieved in 
graphene~\cite{Wangetal:2012,Wangetal:2013,Maoetal:2016,jiang2017tuning,OMJAA:2017}, we analyze the feasibility of experimentally observing the phenomenon of 
geometry-induced wavefunction collapse in graphene. However, our theoretical
prediction of this phenomenon has been made through the solutions of the 
Schr\"{o}dinger equation on a curved surface, so for graphene a critical issue 
is band-gap opening. To carry out the analysis, we first recall some basic 
parameters in our calculation of the LDOS of the Schr\"{o}dinger electron:
the rest mass energy is $Mc^{2}\approx 0.511$MeV and the radial cutoff size 
on a conic surface is $\rho_{0}\approx2\text{\AA}$. In the dimensionless form, 
we have $\sqrt{2ME_{0}}\rho_{0}/\hbar\approx1$ by setting $E_{0}\approx 1$eV. 

As reported in Ref.~\cite{zhou:2007}, bandgap opening can be realized around
$0.26$ eV in graphene through epitaxial growth on the SiC substrate, where the 
gap decreases as the sample thickness increases. It is thus experimentally 
feasible to set the band gap to be $\Delta\approx 0.1$eV, which is related to 
the effective mass of the quasiparticle as $\Delta=M^*v^2_F$. The 
energy-momentum dispersion relation of a massive Dirac Fermion measured from 
a Dirac point in dimensionless form can be written as
\begin{align}
\epsilon=\sqrt{\widetilde{k}^{2}+\widetilde{\Delta}^{2}},
\end{align}
where $\epsilon=\varXi/\varXi_0$, $\widetilde{k}=\hbar v_F k/\varXi_0$ and 
$\widetilde{\Delta}=\Delta/\varXi_0$. Because of the effective mass $M^*$,
it is necessary to transform the original characteristic quantity into a
different form, i.e.,
\begin{align} \label{eq:dimensionless_requirement}
    \sqrt{2ME_{0}}\rho_{0}/\hbar=\sqrt{2M^{*}\varXi_{0}}\xi_{0}/\hbar\approx1
\end{align}
with the new energy unit $\varXi_{0}$ and cutoff radius $\xi_{0}$. In the 
limit of a large gap: $\widetilde{\Delta}\gg\widetilde{k}$, the dispersion 
relation can be approximated as 
\begin{align} \nonumber
\epsilon\approx\widetilde{\Delta}+\delta\epsilon, 
\end{align}
where $\delta\epsilon\equiv\widetilde{k}^{2}/(2\widetilde{\Delta})$. This is
the dispersion relation for a Schr\"{o}dinger particle.

Next, we describe the process required for fabricating a graphene cone and
articulate the possibility of realizing a truncated graphitic cone at the
nanoscale (the setting of our theoretical analysis and computations).
Graphitic cones (or graphene~\cite{LC:2000}) were first reported in 
Ref.~\cite{KDTHLE:1997} in 1997 with the disinclination defects that are 
multiples of $60^\circ$, which correspond to a given number of pentagons:
disk (no pentagons), five types of cones (one to five pentagons), and open
tubes (six pentagons). Another cone-helix structure with a wide distribution
of apex angles in the cone's cross section was experimentally 
realized~\cite{JRDG:2003}. In a very recent experimental 
study~\cite{liu:2022}, spiral graphite cones have been successfully grown 
under extremely harsh conditions such as high temperature and high pressure. 
In addition, open graphitic cones with an apex angle, e.g., $60^\circ$ 
(lampshade structures) were realized~\cite{terrones:2001}. Based on these 
current experimental achievements, we conclude that it is feasible to 
fabricate a nano-truncated graphene cone with an open gap.

According to the dimensionless form Eq.~\eqref{eq:dimensionless_requirement},
the radial cutoff size of a truncated graphite cone can be set as
$\xi_{0}=5\rho_{0}\approx 1$nm, so the characteristic energy is
$\varXi_{0}=2E_{0}=2$eV with the energy band gap $\Delta\approx 0.1$eV.
From our theoretical results, if the graphite cone is shaped as $\alpha=0.5$
(so the apex of the cone is 60$^{\circ}$ as reported in 
Refs.~\cite{KDTHLE:1997,terrones:2001}) and if LDOS is to be detected at the 
radial position $\rho=r\xi_{0}\approx6$nm, where
$r=\exp\left[\pi/(2\alpha)\right]\approx 6$, it would be possible to
observe the collapse oscillations of LDOS with the energy interval
$\varXi=\delta\epsilon\varXi_{0}\in2\times[1,10^{3}]$ $\mu$eV. In this case,
the wavevector is 
\begin{align} \nonumber
\widetilde{k}\approx\sqrt{2\delta\epsilon\widetilde{\Delta}}\approx[10^{-3.5},10^{-2}], 
\end{align}
which is describable by the Schr\"{o}dinger equation, as shown in 
Fig.~\ref{fig:graphene_experimental_scheme}(b).

Our analysis of the experimental feasibility indicates that the phenomenon of 
geometry-induced wavefunction collapse can arise in nanoscale graphene systems,
rendering important to take this phenomenon into consideration when developing
graphene-based devices that involve curved or Riemannian geometry.

\section{Discussion} \label{sec:discussion}

Our study has focused on the quantum states of particles confined on a
truncated conic surface, for which the corresponding geometric potential has
the form of inverse squared distance. 
It has been established for a long time that, semiclassically, this type of 
potential in 3D can cause a particle to collapse to the 
center~\cite{LL:book,nicholson:1962,perelomov:1970,alliluev:1972}. 
The main motivation that we chose to study the conic structure is that
it can be realized in experiments, such as graphite
nanocones~\cite{GS:1994,KDTHLE:1997,ZJW:2003,JRDG:2003} where the issue of
topological phase~\cite{Berry:1988,Anandan:1992,ACW:1997,CLBNGK:2019} was
addressed~\cite{LC:2000,AM:2000,CR:2001,JC:2004,LC:2004,FMC:2008,Akay:2015,PSP:2014}. There were also previous
studies~\cite{FM:2008,PNLM:2010,FSA:2012,ST:2012,DWLKLZ:2016} on the effects of
the geometric potential in terms of the mean and Gaussian
curvatures~\cite{FM:2008,FSA:2012} on the quantum states on the conic surface.
The main contribution of our work is the finding of a class of quantum states
that mimic those arising in atomic collapse, but here the collapse mechanism
is purely geometrical, henceforth the terminology ``geometry-induced
wavefunction collapse.'' In particular, depending on the angular momentum and 
the energy of the particle, the inverse square-distance potential can generate 
bound states, conventional scattering states, and collapse states that are 
essentially an abnormal type of scattering states. The emergence of the 
collapse states was demonstrated through the LDOS that exhibits infinite 
oscillations with the energy near the zero energy point separating the 
scattering states from the bound states. We note that this feature of infinite 
oscillations was previously used to establish the atomic collapse states about 
a Coulomb impurity in graphene~\cite{PNN:2007,SKL:2007b}. From a classical 
point of view, the geometry-induced and Coulomb-impurity induced collapse 
states share a common feature: the particle appears to fall into the center 
but will escape eventually either due to the finite $\rho_0$ or the complex 
eigenenergy. A key difference is that the geometry-induced collapse states 
uncovered here are a nonrelativistic quantum phenomenon while the atomic 
collapse states have a relativistic quantum origin.

The mechanism for the geometry-induced collapse states can be intuitively
understood by noting that the sign of the effective radial potential is 
determined by [Eq.~\eqref{eq:tilde_nu}]
\begin{displaymath}
\frac{l^{2}}{\alpha^{2}}-\frac{1-\alpha^{2}}{4\alpha^{2}},
\end{displaymath}
where the second term is due to the mean curvature of the truncated cone.
For the quantum states corresponding to zero angular momentum, the effective
potential is attractive. The inverse squared distance dependence in
Eq.~\eqref{eq:UG} makes this type of geometry-induced ``Coulomb impurity''
much stronger than a usual Coulomb potential, thereby leading to collapse
states with the classical picture of a particle falling into the center
of the cone. For positive energy states, due to the reflection at $\rho_0$, 
the particle will eventually escape to infinity. More specifically, for
$\alpha\rightarrow 1$, the  geometry induced attractive potential vanishes, the 
quantum states degenerate to those described by the zeroth-order Bessel 
functions, which are scattering states in the 2D plane with a hard hole 
around the center. In this case, neither bound nor collapse states are possible.
In the opposite extreme $\alpha\rightarrow 0$, the depth of the attractive 
potential becomes infinite, so only the bound states are possible. In between 
the two extreme cases where $0<\alpha<1$, collapse states can arise, which is 
promising to be observed in experiments.

\section*{Acknowledgment}

We thank Dr. H.-Y. Xu for discussions during the initial stage of this work.
The work at Arizona State University was supported by the Air Force of
Scientific Research through Grant No.~FA9550-21-1-0186. The work at Lanzhou
University was supported by NSFC under Grants No.~12175090 and No.~12047501.

\appendix

\section{Gaussian and mean curvatures of a conic surface} \label{appendix_A}

From the Gauss-Bonnet theorem
\begin{equation}
	\int\int_{\mbox{int}(\gamma)}KdA=2\pi-\int_{\gamma}\kappa_{g}ds=
2\pi\left(1-\alpha\right),
\end{equation}
where the path $\gamma$ is illustrated in Fig.~\ref{fig:cone_shape}, the
Gaussian curvature satisfies the equation
\begin{equation}
\intop_{0}^{\infty}K\alpha\rho d\rho\int_{0}^{2\pi}d\varphi=2\pi\left(1-\alpha\right).
\end{equation}
The Gaussian curvature of a conic surface is thus given by
\begin{equation} \label{Gaussian curvature}
K=\left(\frac{1-\alpha}{\alpha}\right)\frac{\delta(\rho)}{\rho},
\end{equation}
where the $\delta$-function singularity originates from the apex of the cone.
The mean curvature, the average of the maximal and minimal normal curvatures,
is
\begin{equation}
K_M=\frac{\sqrt{1-\alpha^{2}}}{2\alpha\rho},
\end{equation}
where $k_{1}=1/(\alpha\rho)$,
$k_{1,n}=\sqrt{1-\alpha^{2}}/\left(\alpha\rho\right)$ and $k_2=k_{2,n}=0$.

\section{Solutions of the Schr\"{o}dinger equation in the angular-momentum
representation} \label{appendix_B}

The general solution of Eq.~\eqref{eq:RE} is
\begin{equation}
y(x)=AJ_{\nu}\left(x\right)+BY_{\nu}\left(x\right),
\end{equation}
where the order $\nu$ and $x$ are real or purely imaginary. The series
representation of $J_{\nu}(x)$ is
\begin{equation}
J_{\nu}\left(x\right)=\sum_{k=0}^{\infty}\frac{\left(-1\right)^{k}}
{k!\Gamma(k+\nu+1)}\left(\frac{x}{2}\right)^{2k+\nu},
\end{equation}
which satisfies Eq.~\eqref{eq:RE} regardless of whether the order and the
argument are real or purely imaginary.

For clarity, the quantities $\nu$ and $x$ are defined to be real.
If $\nu$ is real, then $\nu\geq0$; if $\nu$ is purely imaginary, then
write $\nu$ as $i\nu,\nu>0$. Similarly, if $x$ is real, we have $x\geq0$ and if $x$ is purely imaginary, we write $x$ as $ix,x>0$, respectively. Real
$\nu$ values correspond to quantum states of nonzero angular momenta and purely
imaginary $\nu$ values are associated with the zero angular-momentum states.
Real and purely imaginary $x$ values are indicative of positive and negative
energies, respectively. In particular, if the order $\nu$ or the argument $x$
is purely imaginary, $J_{\nu}(x)$ and $Y_{\nu}(x)$ may not be real. Hence,
it is necessary to give some extra definitions for the real Bessel
functions~\cite{dunster1990bessel,olver1997asymptotics}.

If both $\nu$ and $x$ are real, the real solution is
\begin{equation}\label{sol:normal in math}
y\left(x\right)=AJ_{\nu}\left(x\right)+BY_{\nu}\left(x\right).
\end{equation}
If $\nu$ is real but $ix$ is purely imaginary, the real solution is
\begin{equation}
y\left(x\right)=AI_{\nu}\left(x\right)+BK_{\nu}\left(x\right).
\end{equation}
For $i\nu$ purely imaginary and $x$ real, the real solution is
\begin{equation}\label{sol:collapse in math}
y\left(x\right)=AF_{i\nu}\left(x\right)+BG_{i\nu}\left(x\right).
\end{equation}
If both $i\nu$ and $ix$ are purely imaginary, the real solution is
\begin{equation}\label{sol:bound in math}
y\left(x\right)=AL_{i\nu}\left(x\right)+BK_{i\nu}\left(x\right).
\end{equation}
All these solutions can be written as $J_{\nu}\left(x\right)$,
$J_{-\nu}\left(x\right)$, $J_{\nu}\left(ix\right)$, $J_{-\nu}\left(ix\right)$,
$J_{i\nu}\left(x\right)$, $J_{-i\nu}\left(x\right)$, $J_{i\nu}\left(ix\right)$,
$J_{-i\nu}\left(ix\right)$, or their combinations:
\begin{align}
	\label{eq:I}
	I_{\nu}\left(x\right) &=i^{-\nu}J_{\nu}\left(ix\right), \\ \label{eq:K_1}
	K_{\nu}\left(x\right) &=\pi\left[I_{-\nu}\left(x\right)-I_{\nu}\left(x \right)\right]/\left(2\sin\left(\nu\pi\right)\right), \\ \nonumber
    F_{i\nu}\left(x\right) &=\frac{1}{2}\left\{ e^{-\pi\nu/2}H_{i\nu}^{(1)} \left(x\right)+e^{\pi\nu/2}H_{i\nu}^{(2)}\left(x\right)\right\} \\ \label{eq:F}
    &=\frac{1}{2}\left\{ A_{\nu}J_{i\nu}\left(x\right)+iB_{\nu}Y_{i\nu}\left(x\right)\right\} \\ \nonumber
	G_{i\nu}\left(x\right)&=\frac{1}{2i}\left\{ e^{-\pi\nu/2}H_{i\nu}^{(1)}\left(x\right)-e^{\pi\nu/2}H_{i\nu}^{(2)}\left(x\right)\right\} \\ \label{eq:G}
    &=\frac{1}{2i}\left\{ B_{\nu}J_{i\nu}\left(x\right)+iA_{\nu}Y_{i\nu}\left(x\right)\right\} \\ \nonumber
	L_{i\nu}\left(x\right)&=iC_{\nu}\left\{ I_{-i\nu}\left(x\right)+I_{i\nu}\left(x\right)\right\} \\ \label{eq:L}
    &=iC_{\nu}\left\{ i^{i\nu}J_{-i\nu}\left(ix\right)+i^{-i\nu}J_{i\nu}\left(ix\right)\right\} \\ \nonumber
	K_{i\nu}\left(x\right)&=C_{\nu}\left\{ I_{-i\nu}\left(x\right)-I_{i\nu}\left(x\right)\right\} \\ \label{eq:K_2}
    &=C_{\nu}\left\{ i^{i\nu}J_{-i\nu}\left(ix\right)-i^{-i\nu}J_{i\nu}\left(ix\right)\right\},
\end{align}
where
\begin{align}
        \nonumber	
	A_{\nu}&=e^{-\pi\nu/2}+e^{\pi\nu/2}, \\ \nonumber
	B_{\nu}&=e^{-\pi\nu/2}-e^{\pi\nu/2}, \\ \nonumber
	C_{\nu}& =\pi/\left(2\sin\left(i\nu\pi\right)\right).
\end{align}
The power series representations of $F_{i\nu}\left(x\right)$,
$G_{i\nu}\left(x\right)$, $L_{i\nu}\left(x\right)$, and $K_{i\nu}\left(x\right)$
are given by~\cite{dunster1990bessel}
\begin{align}
	F_{i\nu}\left(x\right)& =D_{\nu}\sum_{s=0}^{\infty}\frac{\left(-1\right)^{s}\cos\left(\alpha_{\nu,s}\left(x\right)\right)}{\beta_{\nu,s}}\left(\frac{x}{2}\right)^{2s}, \\
	G_{i\nu}\left(x\right) &=E_{\nu}\sum_{s=0}^{\infty}\frac{\left(-1\right)^{s}\sin\left(\alpha_{\nu,s}\left(x\right)\right)}{\beta_{\nu,s}}\left(\frac{x}{2}\right)^{2s}, \\
	L_{i\nu}\left(x\right) &=M_{\nu}\sum_{s=0}^{\infty}\frac{\cos\left(\alpha_{\nu,s}\left(x\right)\right)}{\beta_{\nu,s}}\left(\frac{x}{2}\right)^{2s}, \\
	K_{i\nu}\left(x\right) &=-M_{\nu}\sum_{s=0}^{\infty}\frac{\sin\left(\alpha_{\nu,s}\left(x\right)\right)}{\beta_{\nu,s}}\left(\frac{x}{2}\right)^{2s},
\end{align}
where
\begin{align}
	\nonumber
	\alpha_{\nu,s}\left(x\right)& =\nu\ln\left(x/2\right)-\phi_{\nu,s}, \\ \nonumber
	\beta_{\nu,s}&=s!\left[\left(\nu^{2}\right)\left(1^{2}+\nu^{2}\right)\cdots\left(s^{2}+\nu^{2}\right)\right]^{1/2}, \\ \nonumber
	D_{\nu} &=\left(\frac{2\nu\tanh\left(\nu\pi/2\right)}{\pi}\right)^{1/2}, \\ \nonumber
	E_{\nu}& =\left(\frac{2\nu\coth\left(\nu\pi/2\right)}{\pi}\right)^{1/2}, \\ \nonumber
	M_{\nu} &=\left(\frac{\nu\pi}{\sinh\left(\nu\pi\right)}\right)^{1/2},
\end{align}
and $\text{ \ensuremath{\phi_{\nu,s}}}=\arg\left\{ \Gamma\left(1+s+i\nu\right)\right\}$, $\phi_{\nu,s}$ is continuous for $0<\nu<\infty$, with
$\lim_{\nu\rightarrow0}\phi_{\nu,s}=0$.

For $x\rightarrow0^{+}$, we have
\begin{align}
	\label{eq:F near zero}
	F_{i\nu}\left(x\right)&\rightarrow D_{\nu}\cos\left(\nu\ln\left(x/2\right)-\phi_{\nu,0}\right)/\nu, \\ \label{G diverge near zero energy}
	G_{i\nu}\left(x\right) &\rightarrow E_{\nu}\sin\left(\nu\ln\left(x/2\right)-\phi_{\nu,0}\right)/\nu, \\
	L_{i\nu}\left(x\right)&\rightarrow M_{\nu}\cos\left(\nu\ln\left(x/2\right)-\phi_{\nu,0}\right)/\nu, \\
	K_{i\nu}\left(x\right)&\rightarrow-M_{\nu}\sin\left(\nu\ln\left(x/2\right)-\phi_{\nu,0}\right)/\nu.
\end{align}
For $x\rightarrow +\infty$, we have
\begin{align}
	\label{eq:F infinity}
	F_{i\nu}\left(x\right)&\rightarrow J_{0}\left(x\right), \\ \label{eq:G infinity}
	G_{i\nu}\left(x\right)& \rightarrow J_{1}\left(x\right), \\ \label{eq:K infinity}
	K_{i\nu}\left(x\right)&\sim\left(\frac{\pi}{2x}\right)^{1/2}e^{-x}, \\ \label{eq:L infinite}
	L_{i\nu}\left(x\right)&\sim\frac{1}{\sinh\left(\nu\pi\right)}\left(\frac{\pi}{2x}\right)^{1/2}e^{x}.
\end{align}
For $\nu\rightarrow 0^{+}$ and any definitive $x$, we have which is consistent with the statement after the work's Eq. (3.3)~\cite{dunster1990bessel}
\begin{align}\label{F converge J_0}
F_{i\nu}\left(x\right)\sim\lim_{\nu\rightarrow0^{+}}\sum_{s=0}^{\infty}J_{0}
\left(s,x\right)\cos\left(\ensuremath{\phi_{\nu,s}}\right)=J_{0}\left(x\right),\\
\lim_{\nu\rightarrow0^{+}}G_{i\nu}(x)\rightarrow Y_{0}(x)(check\; numerically)
\end{align}
where
\begin{align}
    J_{\nu}\left(x\right)&\equiv\sum_{s=0}^{\infty}J_{\nu}\left(s,x\right)
    =\sum_{s=0}^{\infty}\frac{\left(-1\right)^{s}}{s!\Gamma(s+\nu+1)}\left(\frac{x}{2}\right)^{2s+\nu}.
\end{align}
For $x\rightarrow 0^{+}$ and $\nu\rightarrow 0^{+}$, the amplitudes of
$F_{i\nu}$ and $G_{i\nu}$ tend to one and infinity, respectively. In this case,
the function $G_{i\nu}$ can be neglected.

For $x\rightarrow+\infty$, we have
\begin{align}
	F_{i\nu}\left(x\right)&\rightarrow\left(\frac{2}{\pi x}\right)^{1/2}\left\{ \zeta\left(i\nu\right)\cos\alpha-\eta\left(i\nu\right)\sin\alpha\right\}, \\
	G_{i\nu}\left(x\right)&\rightarrow\left(\frac{2}{\pi x}\right)^{1/2}\left\{ \zeta\left(i\nu\right)\sin\alpha+\eta\left(i\nu\right)\cos\alpha\right\}, \\
	J_{\nu}\left(x\right)&\rightarrow\left(\frac{2}{\pi x}\right)^{1/2}\cos\beta,
\end{align}
where $\alpha\equiv x-\pi/4$, $\beta=\alpha-\nu\pi/2$, and for $x\rightarrow+\infty$
\begin{align}
	\nonumber
\zeta\left(i\nu\right)&\equiv\sum_{s=0}^{\infty}\left(-1\right)^{s}\frac{A_{2s}\left(i\nu\right)}{x^{2s}}\rightarrow1 \\ \nonumber
	\eta\left(i\nu\right) &\equiv\sum_{s=0}^{\infty}\left(-1\right)^{s}\frac{A_{2s+1}\left(i\nu\right)}{x^{2s+1}}\rightarrow 0, \\ \nonumber
	A_{s}\left(i\nu\right)&=\frac{\left(4(i\nu)^{2}-1^{2}\right)\cdots\left(-4(i\nu)^{2}-\left(2s-1\right)^{2}\right)}{s!8^{s}},
\end{align}
which lead to Eqs.~\eqref{eq:F infinity} and \eqref{eq:G infinity}.

\bibliography{GIWFC}
\end{document}